%% file: Sr3Fe2O7_final.tex
\documentclass[twocolumn,superscriptaddress,reprint,aps,prl,amsmath,amssymb,floatfix]{revtex4-2}

\usepackage{graphicx}
\usepackage{hyperref}
\hypersetup{pdftitle={Hidden Charge Order in Sr3Fe2O7},pdfauthor={Darren C. Peets},pdfsubject={Strongly correlated electron physics},pdfdisplaydoctitle}
\def\SFO{Sr$_3$Fe$_2$O$_7$}

\begin{document}

\title{Hidden charge order in an iron oxide square-lattice compound}

\author{Jung-Hwa Kim}
\altaffiliation{These authors contributed equally to this work.}
\affiliation{Max-Planck-Institut f\"ur Festk\"orperforschung, D-70569 Stuttgart, Germany}

\author{Darren C.\ Peets}
\altaffiliation{These authors contributed equally to this work.}
%\email{darren.peets@tu-dresden.de}
\affiliation{Max-Planck-Institut f\"ur Festk\"orperforschung, D-70569 Stuttgart, Germany}
\affiliation{Ningbo Institute for Materials Technology and Engineering, Chinese Academy of Sciences, Zhenhai, Ningbo, 315201 Zhejiang, China}
\affiliation{Institut f\"ur Festk\"orper- und Materialphysik, Technische Universit\"at Dresden, D-01069 Dresden, Germany}

\author{Manfred Reehuis}
\affiliation{Helmholtz-Zentrum Berlin f\"ur Materialien und Energie, D-14109 Berlin, Germany}

\author{Peter Adler}
\affiliation{Max-Planck-Institut f\"ur Chemische Physik fester Stoffe, D-01187 Dresden, Germany}

\author{Andrey Maljuk}
\affiliation{Max-Planck-Institut f\"ur Festk\"orperforschung, D-70569 Stuttgart, Germany}
\affiliation{Leibniz Institut f\"ur Festk\"orper- und Werkstoffforschung, D-01171 Dresden, Germany}

\author{Tobias Ritschel}
\author{Morgan C.\ Allison}
\affiliation{Institut f\"ur Festk\"orper- und Materialphysik, Technische Universit\"at Dresden, D-01069 Dresden, Germany}
\author{Jochen Geck}
\affiliation{Institut f\"ur Festk\"orper- und Materialphysik, Technische Universit\"at Dresden, D-01069 Dresden, Germany}
\affiliation{W\"urzburg-Dresden Cluster of Excellence ct.qmat, Technische Universit\"at Dresden, 01062 Dresden, Germany}

\author{Jose R.\ L.\ Mardegan}
\author{Pablo J.\ Bereciartua Perez}
\author{Sonia Francoual}
\affiliation{Deutsches Elektronen-Synchrotron DESY, Hamburg 22603, Germany}

\author{Andrew C.\ Walters}
\affiliation{Max-Planck-Institut f\"ur Festk\"orperforschung, D-70569 Stuttgart, Germany}
\affiliation{Diamond Light Source, Harwell Campus, Didcot OX11 0DE, United Kingdom}

\author{Thomas Keller}
\affiliation{Max-Planck-Institut f\"ur Festk\"orperforschung, D-70569 Stuttgart, Germany}
\affiliation{Max Planck Society Outstation at the Heinz Maier-Leibnitz Zentrum (MLZ), D-85748 Garching, Germany}

\author{Paula M.\ Abdala}
\affiliation{SNBL at ESRF, BP 220, F-38042 Grenoble Cedex 9, France}

\author{Philip Pattison}
\affiliation{SNBL at ESRF, BP 220, F-38042 Grenoble Cedex 9, France}
\affiliation{Laboratory for Quantum Magnetism, \'Ecole polytechnique f\'ed\'erale de Lausanne (EPFL), BSP-Dorigny, CH-1015 Lausanne, Switzerland}

\author{Pinder Dosanjh}
\affiliation{Department of Physics and Astronomy, University of British Columbia, Vancouver, BC, V6T 1Z1 Canada}

\author{Bernhard Keimer}
\email{b.keimer@fkf.mpg.de}
\affiliation{Max-Planck-Institut f\"ur Festk\"orperforschung, D-70569 Stuttgart, Germany}

\begin{abstract}
Since the discovery of charge disproportionation in the FeO$_2$ square-lattice compound Sr$_3$Fe$_2$O$_7$ by M\"ossbauer spectroscopy more than fifty years ago, the spatial ordering pattern of the disproportionated charges has remained ``hidden'' to conventional diffraction probes, despite numerous x-ray and neutron scattering studies. We have used neutron Larmor diffraction and Fe $K$-edge resonant x-ray scattering to demonstrate checkerboard charge order in the FeO$_2$ planes that vanishes at a sharp second-order phase transition upon heating above 332 K. Stacking disorder of the checkerboard pattern due to frustrated interlayer interactions broadens the corresponding superstructure reflections and greatly reduces their amplitude, thus explaining the difficulty to detect them by conventional probes. We discuss implications of these findings for research on ``hidden order'' in other materials.
\end{abstract}

\maketitle

The term ``hidden order'' was coined for $d$- and $f$-electron compounds that undergo a thermodynamic phase transition whose order parameter cannot be identified using conventional experimental methods  \cite{Mydosh2011,Mydosh2020,Cameron2016,Cao2018}. The most prominent example is URu$_2$Si$_2$ whose hidden-order phase has confounded researchers for decades, despite numerous experimental and theoretical studies \cite{Mydosh2011,Mydosh2020}. A lesser known, but equally puzzling case of hidden order has been found in the FeO$_2$ square-lattice compound Sr$_3$Fe$_2$O$_7$, whose electronically active Fe sites are accessible to M\"ossbauer spectroscopy. As early as 1966, \cite{Gallagher1966,Dann1993} M\"ossbauer experiments on slightly oxygen deficient Sr$_3$Fe$_2$O$_7$ revealed a disproportionation of the Fe$^{4+}$ ions into nominal Fe$^{3+}$ and Fe$^{5+}$ valence states around room temperature. Numerous studies since then have confirmed a sharp phase transition at $T_\text{CO} = 340\pm10$\,K in stoichiometric Sr$_3$Fe$_2$O$_7$, but no hints of a charge-ordering transition (such as a symmetry reduction or crystallographic site splitting) have ever been identified in diffraction data \cite{Dann1993,Adler1997,Kobayashi1997,Adler1999,Mori1999,Kuzushita2000,Peets2013}. We have combined two advanced experimental methods, neutron Larmor diffraction (NLD) \cite{Rekveldt2001,Keller2002} and resonant elastic x-ray scattering (REXS) \cite{Lorenzo2012,Fink2013} at the Fe $K$-absorption edge, to resolve this long-standing conundrum. Specifically, we demonstrate checkerboard charge order in the FeO$_2$ layers and show that the ``invisibility'' of charge ordering in Sr$_3$Fe$_2$O$_7$ originates from frustration of the interactions between neighboring layers.

The impact of geometrical frustration on charge order has been widely investigated, beginning with the classical Verwey transition in magnetite (Fe$_3$O$_4$) \cite{Walz2002,Attfield2006}, and has recently been discussed for widely different solids ranging from metal oxides \cite{Ishihara2010,Ikeda2015,Jiang2014} to organic conductors \cite{CanoCortes2011,Oike2015}, and in diverse contexts such as electronic ferroelectricity \cite{Ishihara2010,Ikeda2015}, superconductivity \cite{Jiang2014}, quantum criticality \cite{CanoCortes2011}, and phase-change memory applications \cite{Oike2015}. 

\begin{figure}[htb]
\includegraphics[width=\columnwidth]{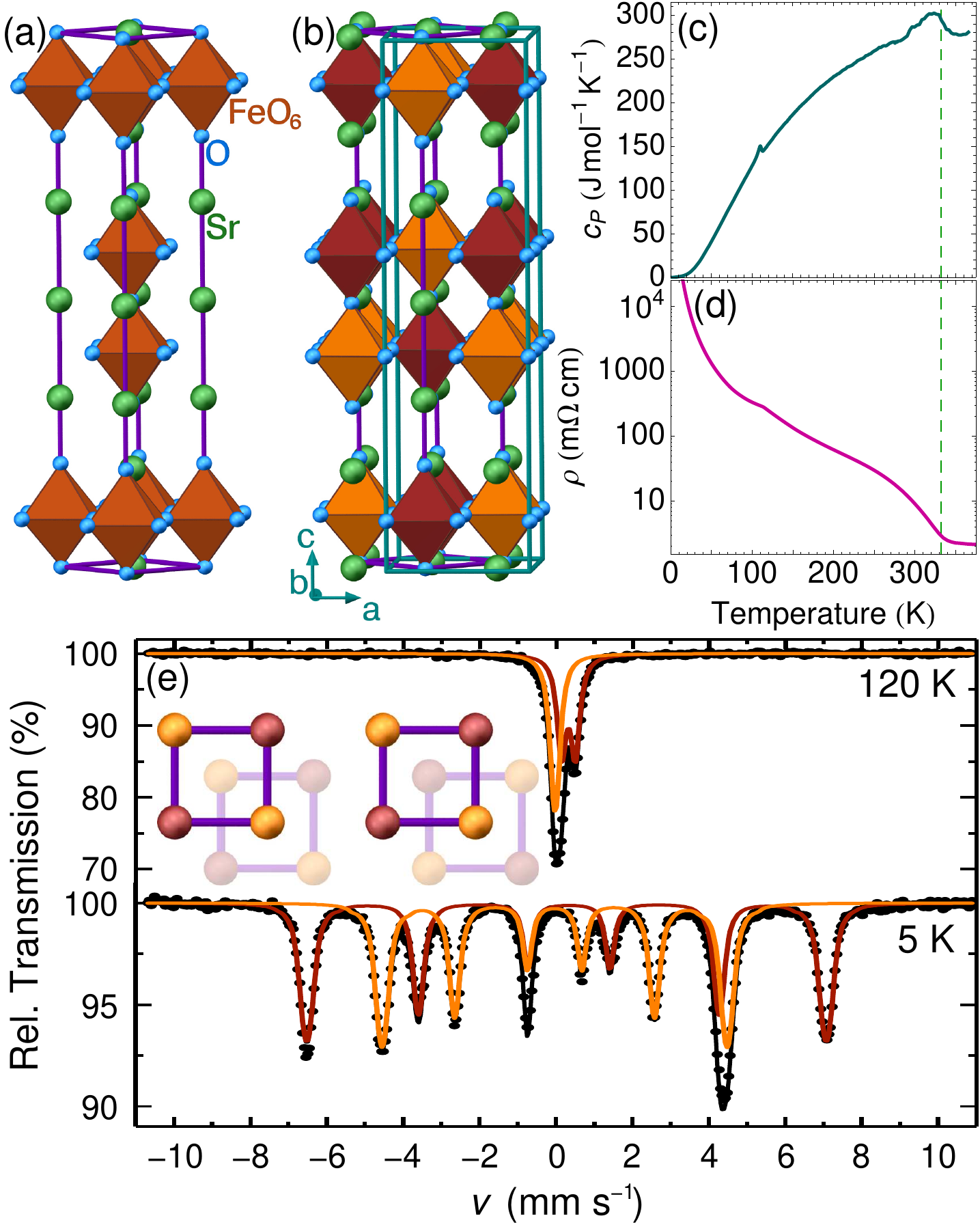}
\caption{\label{structure} Schematic crystal structures of (a) charge-disordered metallic and (b) charge-ordered insulating  Sr$_3$Fe$_2$O$_7$; colors indicate the Fe valence states. (c) Specific heat. An entropy-conserving construction identifies a transition at $T_\text{CO} = 332$ K (dashed line), consistent with transport and diffraction data. (d) In-plane resistivity showing a metal-insulator transition at $T_\text{CO}$. The anomalies at $T \sim 115$ K in panels c and d are due to the onset of helical magnetic order. (e) M\"ossbauer spectra of Sr$_3$Fe$_2$O$_7$ in the paramagnetic and magnetically-ordered phases. The outer and inner components correspond to Fe$^{3+}$- and Fe$^{5+}$-like sites, respectively \cite{Dann1993,Adler1997,Kobayashi1997,Adler1999,KimPRL2019Supp}. Inset: Two degenerate stacking patterns of Fe$^{3+}$- and Fe$^{5+}$-like sites in adjacent bilayers.}
\end{figure}

Sr$_3$Fe$_2$O$_7$ can serve as a model compound for frustrated charge order, because it is chemically stoichiometric and crystallizes in a body-centered tetragonal structure with FeO$_2$ square-lattice bilayers (Fig.~\ref{structure}a). This lattice architecture is common to many materials including Sr$_3$Ru$_2$O$_7$, La$_{2-2x}$Sr$_{1+2x}$Mn$_2$O$_7$, and La$_{2-x}$Sr$_x$CaCu$_2$O$_{6+\delta}$, which have been intensely studied in relation to quantum criticality \cite{Borzi2007}, magnetoresistance \cite{Kimura1996}, and superconductivity \cite{Cava1990}. Helical magnetic order due to competing exchange interactions between the Fe ions sets in at a much lower temperature ($T_\text{N} = 115$\,K) and does not affect $T_\text{CO}$ \cite{Kim2014}.  Our crystallographic data imply that the cooperative Jahn-Teller effect is inactive and orbital order is absent in Sr$_3$Fe$_2$O$_7$, in contrast to isoelectronic manganates such as LaMnO$_3$, but similar to several rare-earth nickelates $R$NiO$_3$\,\cite{Mazin2007}.

Frustration of the Coulomb interactions among valence electrons in Sr$_3$Fe$_2$O$_7$ is caused by the body-centered stacking of FeO$_2$ bilayers, with each Fe located directly above or below the center of a square iron-oxide plaquette in the adjacent bilayer (Fig.~\ref{structure}b and e (inset)). We have found superstructure reflections indicative of checkerboard charge order in the FeO$_2$ layers and demonstrate that stacking disorder due to frustrated interlayer coupling suppresses their amplitude below the detection limit of standard crystalllographic probes. Interlayer frustration thus holds the key to the hidden-order conundrum in Sr$_3$Fe$_2$O$_7$. Possible implications for other hidden-order materials are discussed.

% 332K, 2.69118 J/molK
% 111.6K, 0.686239 J/molK

High-quality single crystals of Sr$_3$Fe$_2$O$_7$ were grown by the floating-zone technique \cite{Peets2013,Maljuk2004}. In order to obtain full oxygen stoichiometry, single-crystalline rods were annealed under 5-6\,kbar of oxygen pressure \cite{KimPRL2019Supp}. Specific heat data (Fig.~\ref{structure}c) demonstrate a second-order phase transition with a sizeable entropy release of $\sim$2.7\,J/mol\,K at $T_\text{CO} = 332$ K. The transition is associated with a strong upturn in the in-plane resistivity (Fig.~\ref{structure}d), in agreement with previous reports \cite{Kuzushita2000,Peets2013}. To confirm charge ordering in our Sr$_3$Fe$_2$O$_7$ samples, we conducted M\"ossbauer experiments on powdered crystals. The resulting spectra (Fig.\ \ref{structure}e) reveal two components in the paramagnetic as well as in the magnetically-ordered phases, indicating charge disproportionation of Fe$^{4+}$ into Fe$^{3+}$- and Fe$^{5+}$-like sites below $T_\text{CO}$. (Note, however, that the high formal charge in Fe$^{4+}$ compounds has to be understood in terms of negative charge transfer energy states, in which the excess holes and the electronic density modulation reside predominantly on the oxygen ligands \cite{Bocquet1992,Green2016}.) The area ratio of 1:1 between the two subspectra confirms full oxidation of the sample and also indicates that the single iron site in the $I4/mmm$ space group has split into two distinct sites with equal population, in agreement with prior work \cite{Gallagher1966,Dann1993,Adler1997,Kobayashi1997,Adler1999,Kuzushita2000}.    

Before addressing the charge-ordered phase, we used single-crystal neutron diffraction to verify the high-temperature crystal
structure from which this order develops.  Refinements \cite{KimPRL2019Supp} in the space group $I4/mmm$ (Fig.~\ref{structure}a) showed no indications of any reduction in symmetry, in agreement with previous work \cite{Dann1993,Mori1999}. The oxygen site bridging two adjacent FeO$_2$ layers refined to full occupancy as expected for stoichiometric Sr$_3$Fe$_2$O$_7$.

\begin{figure}[htb]
\includegraphics[width=\columnwidth]{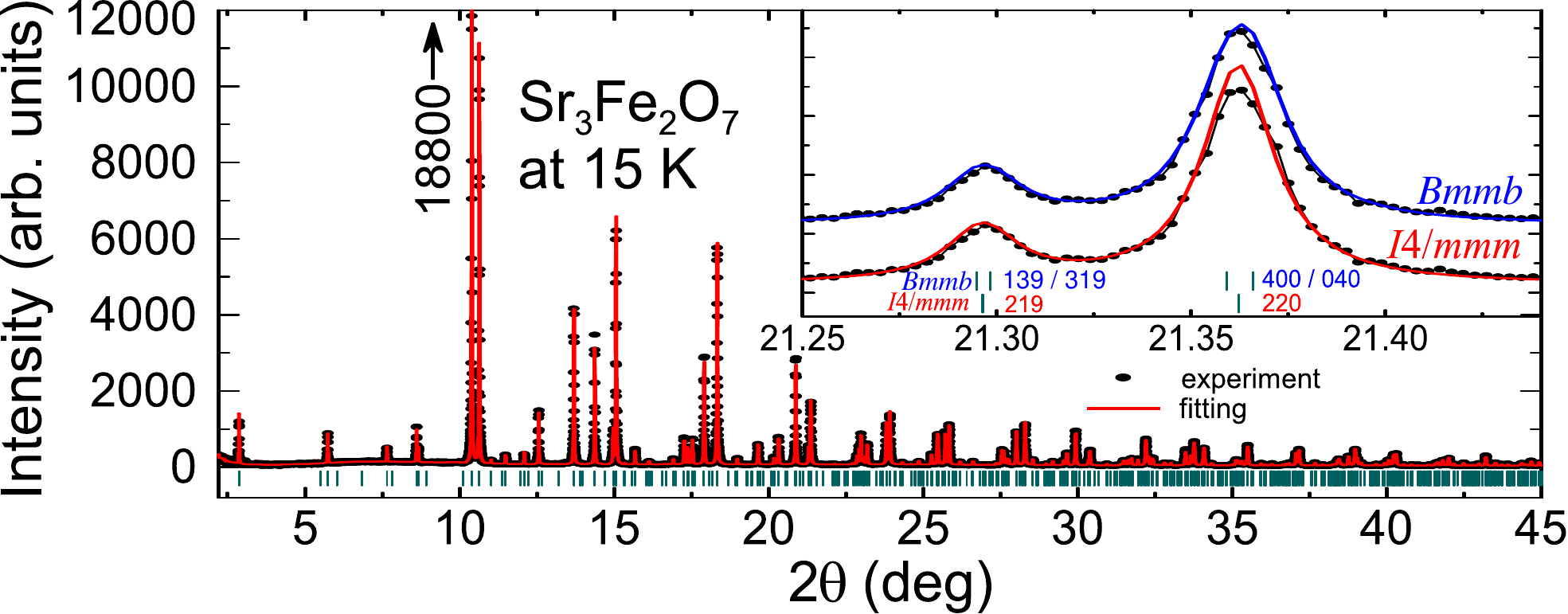}
\caption{\label{synchrotron}High-resolution synchrotron x-ray powder diffraction pattern of the charge-ordered phase at $T = 15$\,K. Inset shows the tetragonal  (2\,1\,9) and (2\,2\,0) Bragg peaks, together with the results of refinements in the $I4/mmm$ and $Bmmb$ space groups.}
\end{figure}

We now turn to the crystal structure for $T < T_\text{CO}$. We first note that neither our neutron diffraction data nor our high-resolution synchrotron x-ray powder pattern (Fig.~\ref{synchrotron}) contained any additional primitive Bragg reflections to indicate a violation of the body-centering condition, nor any obvious splitting of peaks, in agreement with prior work that failed to detect any crystallographic signature of charge disproprotionation \cite{Dann1993,Mori1999}. There was a slight discrepancy between the synchrotron x-ray data and the $I4/mmm$ refinements at a handful of peaks (inset of Fig.~\ref{synchrotron}). To check whether anisotropic strain below $T_\text{CO}$ could explain the peak profile broadening, we used a strain model (Laue class $4/mmm$) in the refinement, but the fit did not improve substantially. Nonetheless,  these deviations alone were not compelling evidence for a change in crystal symmetry. 

\begin{figure}
\includegraphics[width=\columnwidth]{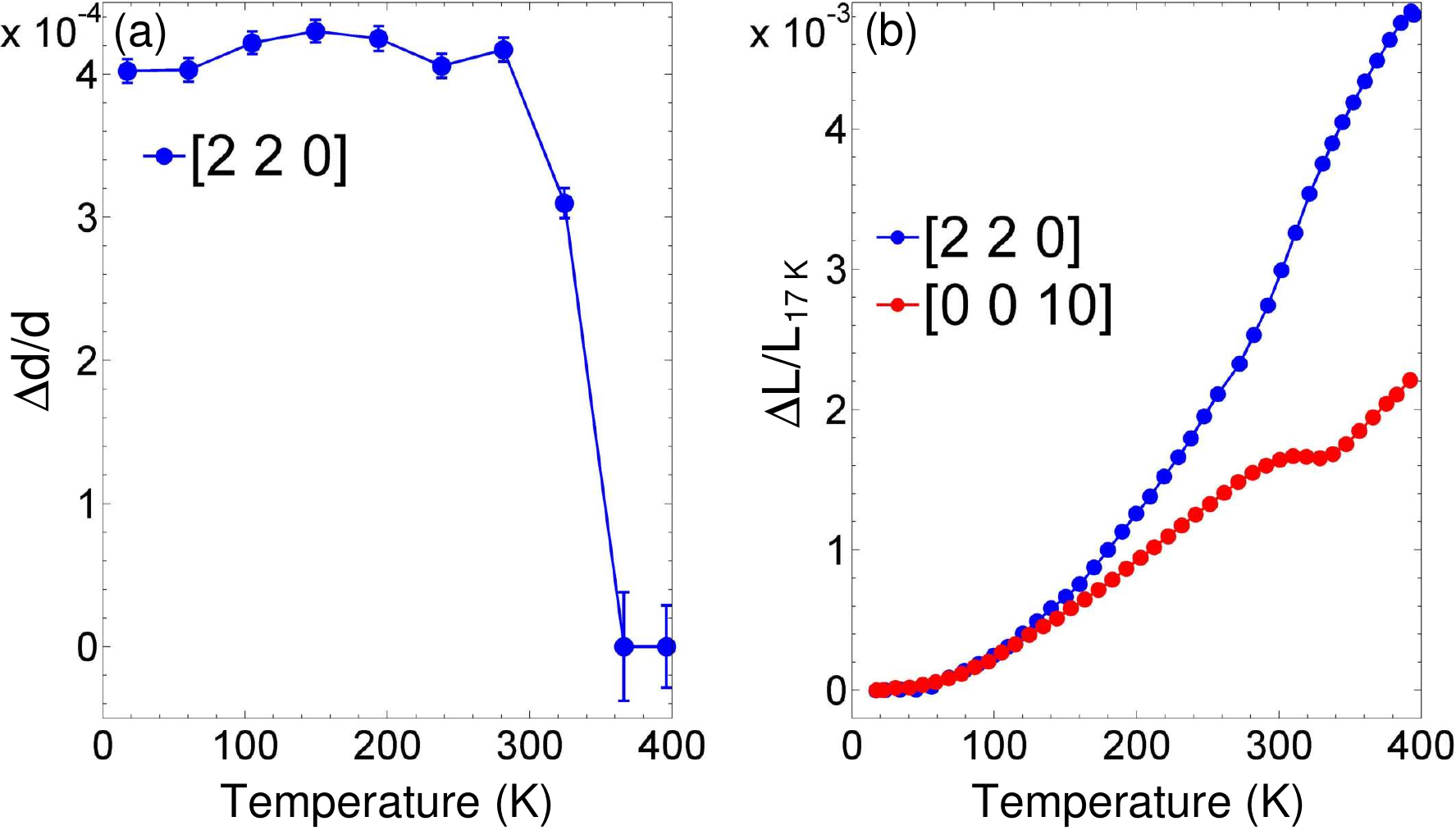}
\caption{\label{Larmor}Neutron Larmor diffraction results. (a) Splitting of the in-plane lattice parameters, $\Delta d/d = (b-a)/a$, extracted from the tetragonal (2\,2\,0) Bragg peak as a function of temperature $T$. In the analysis, the intrinsic peak width of $\Delta Q /Q = 4.8 \times 10^{-4}$ at 400 K was kept constant at all $T$. (b) Thermal expansion extracted from the cumulative Larmor phase of the tetragonal (2\,2\,0) and (0\,0\,10) Bragg peaks, relative to the phase at $T=17$ K.}
\end{figure}

We therefore employed neutron Larmor diffraction on the TRISP spectrometer \cite{TRISP2015} at the Maier-Leibnitz-Zentrum in Garching, Germany. NLD is capable of detecting lattice parameters $d$ and their spread $\Delta d/d$ with a resolution better than $1 \times 10^{-4}$, independent of beam collimation and monochromaticity and of the crystal's mosaic spread \cite{Rekveldt2001,Keller2002,KimPRL2019Supp}. 
Figure \ref{Larmor}a shows that $\Delta d/d$ of the tetragonal (2\,2\,0) Bragg reflection extracted from NLD increases sharply but continuously upon cooling below $T_\text{CO}$, then saturates at a value of $4 \times 10^{-4}$, as expected for the order parameter of a structural phase transition. No comparable change of $\Delta d/d$ is observed along the $c$-axis (not shown), but the thermal expansion of the $c$-axis parameter extracted from the Larmor phase of the (0\,0\,10) reflection provides additional evidence of a continuous structural phase transition at $T_\text{CO}$ (Fig.~\ref{Larmor}b). 

The spread of the in-plane lattice parameters below $T_\text{CO}$ revealed by NLD (Fig.~\ref{Larmor}a) and the slight discrepancy between the synchrotron x-ray data and $I4/mmm$ refinements (Fig.~\ref{synchrotron}) provide clues to the lattice structure in the charge-ordered state. Based on the observation of specific superstructure reflections indicative of a unit cell with doubled in-plane area (see below), we identify the orthorhombic space group $Bmmb$ (an alternate setting of $Cmcm$, No.\ 63) as the simplest crystallographic description compatible with our experimental data. Here, Fe$^{3+}$- and Fe$^{5+}$-like sites in the FeO$_2$ planes alternate in a checkerboard pattern within the plane and also within a bilayer unit along the $c$-axis [Fig.~\ref{structure}(b)]. The checkerboard pattern is analogous to the charge ordering patterns in the pseudocubic perovskite CaFeO$_3$, where Fe$^{3+}$- and Fe$^{5+}$-like sites alternate in all three directions \cite{Woodward2000}.  

Rietveld refinements in $Bmmb$ produced lattice constants $a = 5.43050(3)$\,\AA, $b = 5.43287(3)$\,\AA, and $c = 20.12134(6)$\,\AA. We note that the difference between $a$ and $b$, $(b-a)/a = 4.0 \times 10^{-4}$, is in quantitative agreement with the independent neutron Larmor diffraction results for the in-plane $\Delta d/d$ (Fig.~\ref{Larmor}a).
Complete tables of the resulting structural parameters are given in the Supplemental Materials \cite{KimPRL2019Supp}. The refinement does not indicate any rotations of the FeO$_6$ octahedra such as those observed in nearly-isostructural Sr$_3$Ru$_2$O$_7$ \cite{Shaked2000,Kiyanagi2004} and in CaFeO$_3$.  \cite{Woodward2000} Since substantial rotations and distortions of the FeO$_6$ octahedra are present in CaFeO$_3$ even above $T_\text{CO}$, the splitting of its orthorhombic Bragg reflections, $\Delta Q/Q$, is $\sim$20$\times$ that of the tetragonal peaks in Sr$_3$Fe$_2$O$_7$ for $T < T_\text{CO}$. Charge order then manifests as a spectral-weight shift between the split peaks, which is readily resolved by standard diffraction probes \cite{Woodward2000}.

To search for weak superstructure reflections that are allowed in $Bmmb$ but not in $I4/mmm$, we first investigated a $\sim$10\,$\mu$m-diameter Sr$_3$Fe$_2$O$_7$ single crystal at the BM01A beamline at the ESRF, using the wavelength 0.6973\ \AA. In $Bmmb$, such reflections occur at $(h,0,l)$ for $h + k = 2n$ and $(0,k,l)$ for $l = 2n$. In $I4/mmm$, these have non-integer indices $(h/2, k/2, l)$ and are therefore forbidden. We found no intensity at the position of any superstructure reflection, and were able to place an upper bound of 10$^{-5}$ on the ratio of the peak intensities $I$($h$\,$k$\,$l$)\,/\,[$I$(1\,1\,5) + $I$($-$1\,1\,5)] (in the orthorhombic setting), which should be of order $ 10^{-4}$ according to our refinement \cite{KimPRL2019Supp}. If our space group assignment is correct, this finding implies that the superstructure reflections are broadened by disorder so that their amplitude is reduced below the detection limit.

\begin{figure}[htb]
  \includegraphics[width=\columnwidth]{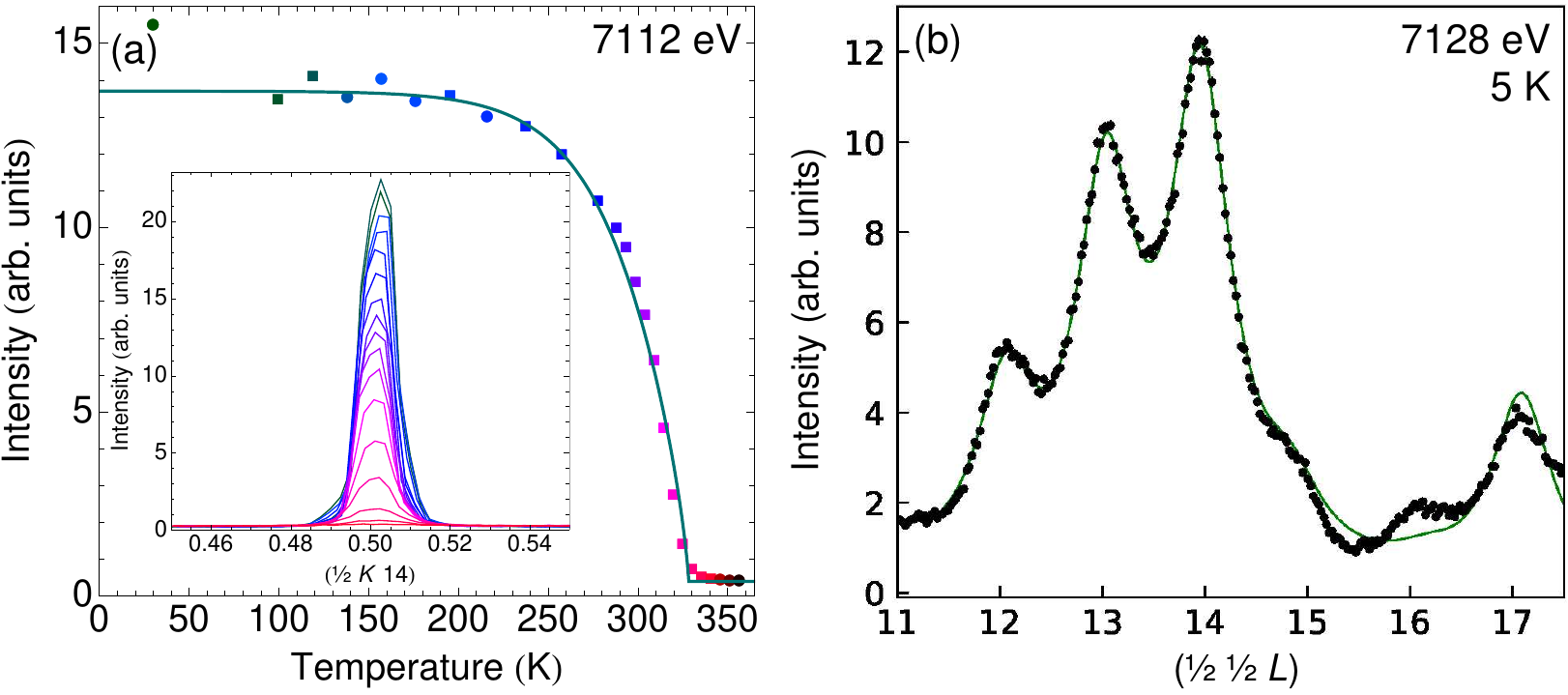}
  \caption{\label{RXD}(a) Temperature dependence of the integrated intensity of the $(\frac12\,\frac12\,14)$ superstructure reflection measured by Fe $K$-edge REXS slightly off resonance (photon energy 7112 eV). The line is a guide to the eye. Inset: In-plane $K$-scans at temperatures marked by squares in the main panel, demonstrating the absence of $T$-dependent shifts or broadening of this reflection.  (b) $L$-scan on resonance at $T = 30$ K. The line is the result of a calculation that considers only the Fe sites, assuming differing charge. The Fe positions were taken from the crystallographic refinement \cite{KimPRL2019Supp}, and only the width, overall intensity, and the imbalance between the population of orthorhombic twin domains ($45:55$\%) were fitted. Reciprocal-space locations refer to the high-temperature $I4/mmm$ cell. $(\frac12\,\frac12\,14)$ in $I4/mmm$ is equivalent to (1\,0\,14)/(0\,1\,14) in $Bmmb$.}
\end{figure}

To enhance the sensitivity to the diffraction signal from a charge modulation on the Fe sites, we performed single-crystal REXS \cite{Lorenzo2012,Fink2013} measurements at the Fe $K$-edge (photon energy 7128 eV) at beamline P09 at PETRA-III, DESY, in Hamburg, Germany~\cite{Strempfer2013}.  The incoming polarization was perpendicular to the scattering plane, and the outgoing polarization was not analyzed. As shown in Fig.~\ref{RXD}, REXS indeed enables the detection of superstructure peaks at the positions predicted for the space group $Bmmb$.  The $(\frac{1}{2},\frac{1}{2},l)$ reflections (in $I4/mmm$ notation) are direct manifestations of the unit-cell doubling due to checkerboard charge order. The intensity of the superstructure reflections decreases continuously with increasing $T$ and vanishes at $T = T_\text{CO}$ (Fig.~\ref{RXD}a). Figure \ref{RXD} represents our most crucial result, as it demonstrates that the checkerboard ordering pattern is correct, and has temperature dependence consistent with M\"ossbauer (Fig.~\ref{structure}) and neutron Larmor diffraction (Fig.~\ref{Larmor}) data.

In addition to the data shown in Fig.~\ref{RXD}, we surveyed $\sim 50$ reciprocal space positions, including primitive reflections forbidden in $Bmmb$ \cite{KimPRL2019Supp}. In particular, superstructure peaks having a temperature dependence consistent with those shown in Fig. 4 were also found at $(\frac{1}{2},\frac{3}{2},l)$ positions (tetragonal cell) for both even and odd $l$, but not at tetragonal $(0,0,2n+1)$ or $(1,1,2n+1)$, as expected for $Bmmb$.  The structure factor of these reflections is roughly consistent with a model that only considers the contribution of the iron atoms resulting from the structural refinement (Fig.~\ref{RXD}b). In this model, the structure factor of the $(\frac{1}{2},\frac{1}{2},l)$ reflections shown in Fig. 4b is given by $4f_1 \sin(2\pi z_1 l) + 4f_2 \sin(2\pi z_2 l)$, where $f_{1,2}$ are the form factors of inequivalent Fe ions, and $z_1 \approx -z_2 \approx 0.097$ their $c$-axis positions measured from the center of the unit cell (Fig. 1b center) \cite{KimPRL2019Supp}. Some deviations from the model calculations are apparent, possibly indicating a contribution from FeO$_6$ octahedral distortions which modulate the Fe $4p$ intermediate state of $K$-edge REXS, as recently found in experiments on nickel oxides \cite{Lu2016}. Since the positional parameters of the O atoms cannot be accurately extracted from the structural refinement  \cite{KimPRL2019Supp}, comprehensive modelling of the REXS intensity on and off resonance goes beyond the scope of the current paper and will be the subject of future work. 

Here we emphasize that the observation of the superstructure reflections and the corresponding extinction rules completes the space-group assignment in the charge-ordered state. The low-temperature space group has to accommodate iron ions in two different valence states as found by the M\"ossbauer experiments. Since the phase transition at $T_{CO}$ is second order, we considered subgroups of $I4/mmm$ with two crystallographically different Fe sites. The observation of peaks at half-integer positions in $h$ and $k$ requires a unit cell with a doubled in-plane area. This requirement leads to $Fmmm$ and its direct subgroups. Out of these, $Cmme$ and $Cmce$ do not support charge order. $Fmmm$ itself and its subgroups $Fmm2$, $F222$, $Ccce$, and $Cccm$ can also be ruled out, because $(\frac{1}{2},0,l)$ and $(0,\frac{1}{2},l)$ reflections were found for both odd and even $l$ (Fig. 4b). Likewise, the non-observation of (1,1,7) and (1,1,9) (again in $I4/mmm$ notation) excludes $C2/m$. This leaves $Bmmm$ and $Bmmb$, which feature in-plane checkerboards with uniform and alternating stacking within a unit cell (and corresponding structure factors with cosine-like and sine-like dependence on $l$), respectively. The experimentally observed sine-like structure factor (Fig. 4b) singles out $Bmmb$, which is also favored by electrostatic and structural considerations. Deviations from orthorhombic symmetry (such as a monoclinic distortion) were not found outside the experimental uncertainty.

The REXS data yield insight into the origin of the ``invisibility'' of the charge-ordered state to standard diffraction probes. Whereas the width of the superstructure reflections is resolution-limited in the FeO$_2$ planes (which implies a lower bound of $\sim 100$ tetragonal lattice spacings on the in-plane correlation length), the reflections are broadened into diffuse streaks along the $c$-axis. From their momentum width, we infer a domain size of $\sim 0.65$ lattice spacings along $c$. Because of the corresponding reduction of the peak amplitude by about two orders of magnitude, the superstructure peaks are below the detection threshold of standard neutron and non-resonant x-ray scattering. The correlation along the $c$-axis is strong within, but weak between bilayer units, so that the order can be regarded as quasi-two-dimensional. Nonetheless, the observation of well-defined superstructure peaks implies that the space group is correct and the stacking is not random. It is interesting to point out that the Ising symmetry of the charge order parameter allows a finite-temperature phase transition in two dimensions \cite{Onoda2004}, which helps explain the sharp transition in the thermodynamic, transport, and diffraction data even in the presence of substantial stacking disorder. 

The insights gained from our resolution of the long-standing Sr$_3$Fe$_2$O$_7$ conundrum provide interesting perspectives for research on hidden order in other compounds. First, we note that  M\"ossbauer spectroscopy provides a sensitive, direct probe of the charge and spin density modulation in Sr$_3$Fe$_2$O$_7$. Without M\"ossbauer data (which are only obtainable on a small number of compounds with M\"ossbauer-active elements), the origin of the prominent phase transition in Sr$_3$Fe$_2$O$_7$ would have been far less evident. This is the case, for instance, for the layered iridates with reported hidden-order transitions \cite{Cao2018} as well as the archetypical hidden-order compound URu$_2$Si$_2$ \cite{Mydosh2011,Mydosh2020}, which also crystallize in $I4/mmm$ with the same body-centered stacking pattern of electronically active atoms (i.e., iridium or uranium) as in Sr$_3$Fe$_2$O$_7$.  
%A minute splitting of the tetragonal Bragg reflections in the hidden-order phase of URu$_2$Si$_2$ has been analyzed in terms of the orthorhombic space group $Fmmm$ \cite{Tonegawa2014}.  In principle, the mechanism of frustration relief we identified is not limited to monopolar charge order, but may also be active for checkerboard ordering of higher-order multipoles, as discussed for URu$_2$Si$_2$ \cite{Mydosh2011,Mydosh2020}. Superstructure reflections in the diffraction pattern of such an order parameter might then be subject to broadening due to the stacking disorder, in analogy to our data on Sr$_3$Fe$_2$O$_7$. 

We also note that the orthorhombic distortion breaks the four-fold rotational symmetry of the tetragonal host lattice, and is thus expected to induce two-fold angular modulations in thermodynamic and transport quantities if domain averaging can be avoided (by reducing the sample volume or by applying external strain). Such modulations have indeed been identified in experiments by Okazaki and coworkers on URu$_2$Si$_2$ \cite{Okazaki2011}, and were attributed to ``nematic'' order, that is, a state with broken rotational symmetry that maintains the translational symmetry of the host lattice. Our observation of the orthorhombic supercell in Sr$_3$Fe$_2$O$_7$ suggests an alternative explanation of this behavior. The possible sensitivity of the orthorhombic domain size along the $c$-axis to cooling protocols or defects might help explain why the experiment of Okazaki {\it et al.} has been difficult to reproduce \cite{Mydosh2020}. 

In summary, our NLD and REXS experiments have resolved the 50-year old puzzle of ``hidden'' charge order in Sr$_3$Fe$_2$O$_7$. The results highlight the need for further investigations of the influence of frustration and disorder on experimental observables in hidden-order phases of other materials. Finally, we point out that the high ordering temperature of Sr$_3$Fe$_2$O$_7$ might enable device applications akin to those recently proposed for organic compounds with frustrated charge order \cite{Oike2015}.

\begin{acknowledgments}
We thank an anonymous referee for insightful comments on an earlier version of this manuscript. We acknowledge funding from the Deutsche Forschungsgemeinschaft (DFG, German Research Foundation), Collaborative Research Center TRR 80 (project-ID 107745057), and through projects C03 and C06 of the Collaborative Research Center SFB 1143 (project-ID 247310070); the National Natural Science Foundation of China (Grant No.~11674367), and the Zhejiang Provincial Natural Science Foundation (Grant No.~LZ18A040002).  DCP is supported by the Chinese Academy of Sciences through 2018PM0036.  The authors are grateful to the groups of R.\ Dinnebier, R.\ Kremer, and L.\ Schr\"{o}der, and the staff of beamline P02.1 at PETRA-III for experimental support. We thank HZB for the allocation of neutron diffraction beamtime; DESY (Hamburg, Germany), a member of the Helmholtz Association HGF, for resonant diffraction beamtime; the European Synchrotron Radiation Facility for provision of synchrotron facilities and access to beamlines BM01A and BM01B; and the Heinz Maier-Leibnitz Zentrum (MLZ), Garching, Germany, for use of the TRISP spectrometer at FRM II.
\end{acknowledgments}

\bibliography{327_structure_arXiv}

\appendix
\renewcommand{\thefigure}{S\arabic{figure}}
\renewcommand{\thesection}{S\arabic{section}}
\section{--- Appendix:  Supplemental Material ---}

\input{supp_content}

\end{document}

%% file: supp_content.tex
\section{Crystal Growth and Characterization}

High-quality single crystalline rods of Sr$_3$Fe$_2$O$_7$ grown by the floating zone technique \cite{Peets2013,Maljuk2004} were annealed to full oxygen stoichiometry following two different temperature programs:  The material studied here by M\"ossbauer spectroscopy, resistivity, neutron diffraction, Larmor diffraction, and synchrotron powder diffraction was annealed under 6~kbar of oxygen pressure at 550$^\circ$C for 100~h then cooled slowly to room temperature.  Crystals used for resonant x-ray diffraction and synchrotron single-crystal diffraction were annealed for 48 hours at 450$^\circ$C in 5~kbar of oxygen to rapidly oxygenate the sample, cooled to 400$^\circ$C in 6 hours and held there for 24 hours to ensure equilibrium, cooled to 350$^\circ$C in 96 hours then to 275 in 48 hours to maximize the oxygen content, then cooled to room temperature in an additional 24 hours. Specific heat measurements were performed on both sets of samples and were indistinguishable, so these were averaged.  The oxygen contents were verified by thermogravimetry, and by refinement of diffraction data.  Thermogravimetric analysis indicated an oxygen content of 6.92 for the former annealing program and 6.96 for the latter, both within the uncertainty of 7.00, and additionally showed a glitch on warming at 332(5)\,K, consistent with the charge-order transition.

The specific heat was measured on single crystals of Sr$_3$Fe$_2$O$_7$ in a Quantum Design Physical Properties Measurement System (PPMS), in zero field and for fields up to 9\,T along the $c$ axis.  No hysteresis nor field dependence was observed above the magnetic transition, and as mentioned above, measurements on crystals from the two batches were indistinguishable and were averaged.  Samples were attached to the sample holder using Apiezon N grease for measurements below 200\,K;  Apiezon H grease was used for higher-temperature measurements to avoid the glass transition of Apiezon N grease.  Resistivity was measured in standard four-probe geometry in a Quantum Design PPMS. Gold wires were attached with silver epoxy, which was allowed to cure for several hours at 180--200$^\circ$C in air before the crystal was mounted to the sample puck with GE Varnish.  Thermogravimetric analysis had previously indicated that at these temperatures the oxygen mobility remains extremely low, the oxygen content does not change, and any intercalated water tends to deintercalate, ensuring that the resistivity samples were not altered or damaged while curing the epoxy.  

\section{Neutron diffraction}

Single-crystal neutron diffraction was performed on the four-circle  diffractometer E5 at the BER-II reactor (Helmholtz-Zentrum Berlin, Germany)
using the neutron wavelength $0.896$~\AA. Refinements were carried out with the program {\sc Xtal}3.4\,\cite{Hall1995} using the nuclear scattering lengths $b$(O)~=~5.805~fm, $b$(Fe)~=~9.54~fm, and $b$(Sr)~=~7.02~fm \cite{ITC2006}.  A data set of 1302 reflections (303 unique) was collected at 390~K, well above $T_\text{CO}$. Refinements of a total of 19 parameters (i.e., the overall scale and extinction factors, 4 positional parameters, and 13 anisotropic thermal parameters) in the space group $I4/mmm$ showed no indications of any reduction in symmetry.  Data taken below the charge order transition showed no change.

\section{Nonresonant x-ray diffraction}

To determine the crystal structure for $T < T_\text{CO}$, high-resolution synchrotron powder diffraction measurements were performed at the BM01B (Swiss-Norwegian) beamline at the ESRF (Grenoble, France) using the wavelength $\lambda = 0.5035$~\AA. The data were refined using the program {\sc fullprof} \cite{Fullprof2}, using the atomic scattering factors provided therein. However, no peak splittings or additional superstructure reflections were observed. The search for weak superstructure reflections continued at beamline BM01A at the ESRF, using a $\sim$10\,$\mu$m-diameter Sr$_3$Fe$_2$O$_7$ single crystal.  Data were collected with a Pilatus area detector, using the photon wavelength 0.6973\,\AA.

\begin{figure}[htb]
  \includegraphics[width=0.5\columnwidth]{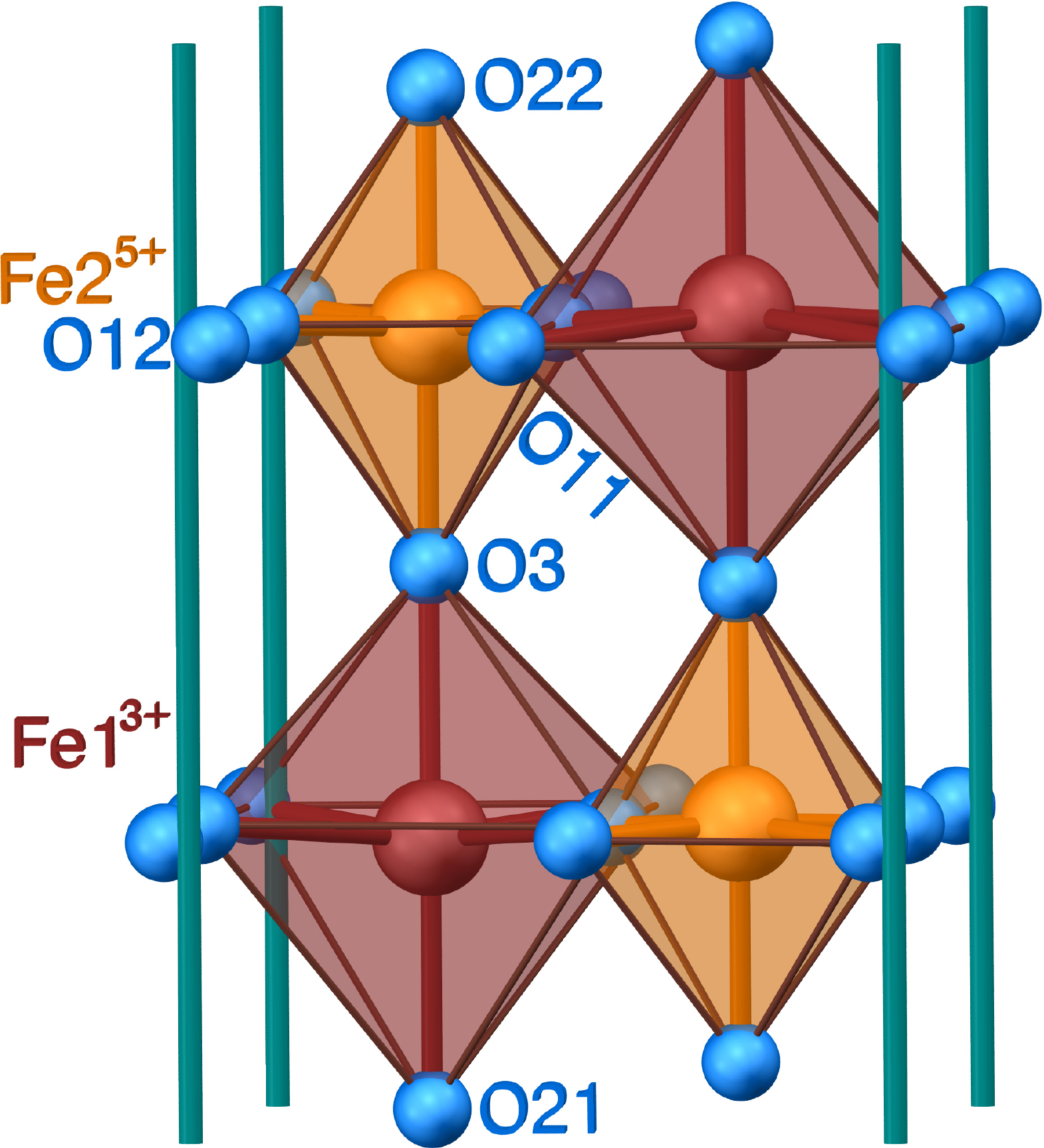}
 \caption{\label{sites} Labelling of the crystallographic sites in the charge-ordered phase. The distortions are exaggerated for clarity.}
\end{figure}

\begin{table}[htb]
  \caption{\label{I4mmm}Results of the crystal structure refinement of \SFO\ from single-crystal neutron diffraction data ($\lambda$ = 0.89\,\AA) collected at 390\,K. The refinement was carried out in the tetragonal space group $I4/mmm$, resulting in a residual $R_F$ = 0.068. The thermal parameters $U_{ij}$ (given in 100\,\AA$^2$) are in the form $\exp\left[-2\pi^2\left(U_{11}h^2(a^*)^2 + ...2U_{13}hla^*c^*\right)\right]$. For symmetry reasons, in $I4/mmm$ the values $U_{11}$ and $U_{22}$ are identical for the atoms Sr1, Sr2, Fe, O1 and O3, and all parameters $U_{12}$, $U_{13}$ and $U_{23}$ are equal to zero.  O1 is equatorial, O2 is apical, and O3 is the shared apical oxygen in the center of the bilayer.}
  \begin{tabular}{lcllr@{.}lr@{.}lr@{.}lr@{.}lr@{.}l}\\ \hline\hline
    & Site & $x$ & $y$ & \multicolumn{2}{c}{$z$} & \multicolumn{2}{c}{$U_{11}$} & \multicolumn{2}{c}{$U_{22}$} & \multicolumn{2}{c}{$U_{33}$} & \multicolumn{2}{c}{Occ} \\ \hline
    Sr1 & 2$b$ & 0 & 0 & \multicolumn{2}{l}{$\frac12$} & 1&32(8) & 1&32 & 1&03(8) & \multicolumn{2}{l}{1}\\
    Sr2 & 4$e$ & 0 & 0 & 0&31656(8) & 1&64(6) & 1&64 & 0&97(6) & \multicolumn{2}{l}{1}\\
    Fe & 4$e$ & 0 & 0 & 0&09745(6) & 0&79(4) & 0&79 & 0&70(5) & \multicolumn{2}{l}{1}\\
    O1 & 8$g$ & 0 & $\frac12$ & 0&09432(8) & 1&52(8) & 1&33(7) & 1&23(5) & \multicolumn{2}{l}{1}\\
    O2 & 4$e$ & 0 & 0 & 0&19309(11) & 1&32(7) & 1&32 & 0&99(6) & \multicolumn{2}{l}{1}\\
    O3 & 2$a$ & 0 & 0 & \multicolumn{2}{l}{0} & 2&07(14) & 2&07 & 1&64(13) & 0&983(23)\\ \hline\hline
  \end{tabular}\\
  $a$ = $b$ = 3.846(4)\,\AA, $c$ = 20.234(2)\,\AA, $V$ = 299.4(6)\,\AA$^3$
\end{table}

\begin{table}[htb]
  \caption{\label{Bmmb}Results of the crystal structure refinements of \SFO\ from high-resolution synchrotron powder diffraction data ($\lambda$ = 0.5035\,\AA) collected at 15\,K. The refinements were carried out in the orthorhombic  space group $Bmmb$. The parameter z(O3) was not allowed to vary, and the thermal parameters $B$ were constrained to be identical for each element.}
  \begin{tabular}{lcr@{.}lr@{.}lr@{.}lr@{.}l}\\ \hline\hline
    $Bmmb$ & Site & \multicolumn{2}{c}{$x$} & \multicolumn{2}{c}{$y$} & \multicolumn{2}{c}{$z$} & \multicolumn{2}{c}{$B$ (\AA$^2$)} \\ \hline
    Sr1 & 4$c$ & \multicolumn{2}{l}{0} & \multicolumn{2}{c}{$\frac34$} & 0&2500(12) & 0&290(4)\\
    Sr21 & 4$c$ & \multicolumn{2}{l}{0} & \multicolumn{2}{c}{$\frac34$} & 0&4330(1) & 0&290\\
    Sr22 & 4$c$ & \multicolumn{2}{l}{0} & \multicolumn{2}{c}{$\frac34$} & 0&0670(1) & 0&290\\
    Fe1 & 4$c$ & \multicolumn{2}{l}{0} & \multicolumn{2}{c}{$\frac34$} & 0&6528(15) & 0&306(10)\\
    Fe2 & 4$c$ & \multicolumn{2}{l}{0} & \multicolumn{2}{c}{$\frac34$} & 0&8471(15) & 0&306\\
    O1 & 16$h$ & 0&248(6) & 0&505(5) & 0&3450(1) & 0&62(3)\\
    O21 & 4$c$ & \multicolumn{2}{l}{0} & \multicolumn{2}{c}{$\frac34$} & 0&5573(24) & 0&62\\
    O22 & 4$c$ & \multicolumn{2}{l}{0} & \multicolumn{2}{c}{$\frac34$} & 0&9445(24) & 0&62\\
    O3 & 4$c$ & \multicolumn{2}{l}{0} & \multicolumn{2}{c}{$\frac34$} & \multicolumn{2}{c}{$\frac34$} & 0&62\\ \hline\hline
    \multicolumn{10}{l}{$a$~=~5.43050(3)\,\AA, $b$~=~5.43287(3)\,\AA, $c$~=~20.12137(7)\,\AA,}\\
  \end{tabular}
\end{table}

\begin{table}[htb]
  \caption{\label{bond}Fe--O bond lengths $d$ in \SFO\ (in \AA) at 390\,K, including the average bond length $d_{av}$.}
  \begin{tabular}{lr@{.}l}\\ \hline\hline
    \multicolumn{3}{c}{\SFO\ at 390\,K, in $I4/mmm$}\\ \hline
    $d$(Fe--O1) $\times4$ (eq) & 1&9244(22)\\
    $d$(Fe--O2) $\times1$ (ap) & 1&9352(26)\\
    $d$(Fe--O3) $\times1$ (br) & 1&9718(13)\\
    $d_{av}$(Fe--O) & 1&9341(21)\\ \hline\hline
  \end{tabular}\\
\end{table}

\section{Crystallographic Parameters}

The atomic coordinates resulting from refinements of neutron and synchrotron x-ray diffraction data at 390 and 15\,K, respectively, are shown in Tables~\ref{I4mmm} and \ref{Bmmb}, and the Fe--O bond lengths obtained at 390\,K are given in Table~\ref{bond}. Figure \ref{sites} indicates the labelling of the crystallographic sites. Somewhat enlarged standard deviations were obtained from synchrotron data for the positional parameters of the oxygen atoms in the low-temperature phase (Table~\ref{Bmmb}). This can be ascribed to the fact that these parameters are highly correlated, and to the fact that the scattering power of the oxygen atoms is relatively weak in x-ray diffraction.

The space group $Bmmb$ generates a general position 16$h$($x,y,z$) for the equatorial O1 atoms. Here the three atomic positions could be individually refined; only the $z$ parameter of O3 was fixed to be $z = 0.75$. Due to the enlarged standard deviations of the positional parameters of the oxygen atoms, it was not possible to determine individual bond lengths $d$(Fe--O) with good accuracy. However, we have obtained reasonable averaged bond lengths $d_{av}$(Fe1) = 1.950\,\AA\ and $d_{av}$(Fe2) = 1.917\,\AA.

\section{Neutron Larmor diffraction}

\begin{figure}[htb]
  \includegraphics[width=0.65\columnwidth]{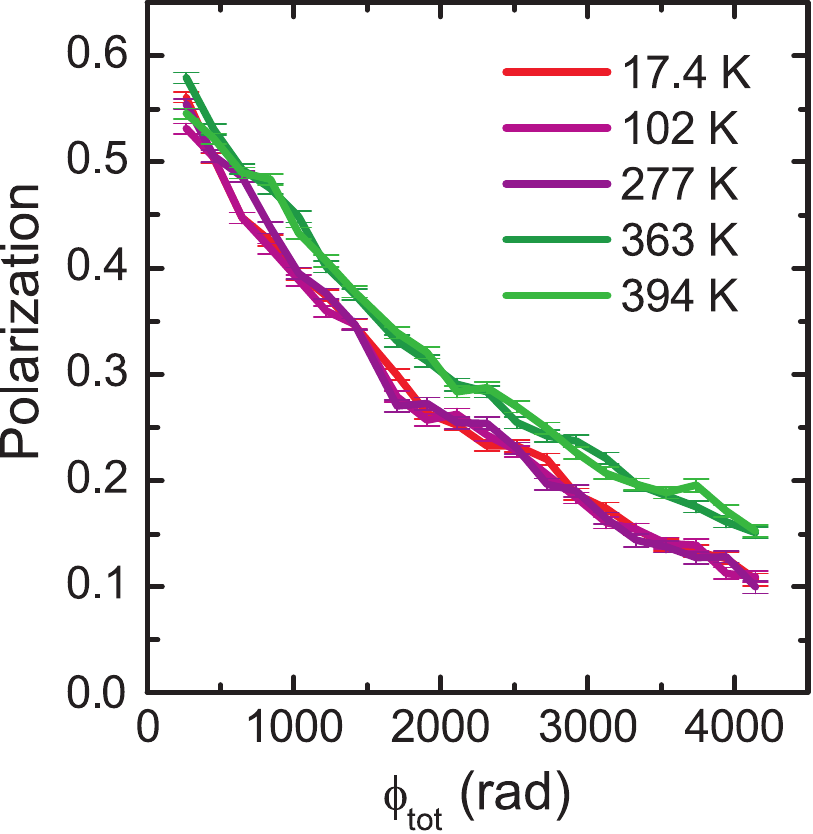}
  \caption{\label{TRISPsupp}Raw polarization data underpinning the neutron Larmor diffraction results on the tetragonal (2\,2\,0) peak.  The increased depolarization at lower temperatures indicates a broadening of the lattice parameter.}
\end{figure}

High-resolution neutron Larmor diffraction data were taken at the resonant spin-echo triple-axis spectrometer TRISP at the FRM-II (Garching, Germany) with neutron wavevector $2.9$ \AA$^{-1}$.
%In this technique, the spin-polarized neutron beam is subjected to identical magnetic fields before and after scattering from the sample. The field strength determines the angle through which the neutron spin direction rotates due to Larmor precession. Because the incoming and outgoing neutron velocities are related by the Bragg condition, the cumulative Larmor phase is directly proportional to the wavevector $Q$ of the Bragg peak being probed, and a spread in this wavevector, $\Delta Q/Q$, induces dephasing of the neutron spins and thus reduces the polarization of the scattered beam. These parameters can be measured through polarization analysis, and can then be converted into $d$ and $\Delta d/d$ through the Bragg condition.
The crystal was aligned in the tetragonal $(HHL)$ scattering plane and cooled in a closed-cycle refrigerator, and the tetragonal (2\,2\,0) and (0\,0\,10) nuclear Bragg peaks were investigated in zero applied field. The outgoing neutron polarization was measured as a function of the magnetic field applied along the incident and scattered neutron paths, parallel to the tetragonal (2\,2\,0) Bragg planes.  The measured polarization data for the (2\,2\,0) peak are shown in Fig.~\ref{TRISPsupp}.

\section{Resonant elastic x-ray scattering}

\begin{figure}[htb]
  \includegraphics[width=\columnwidth]{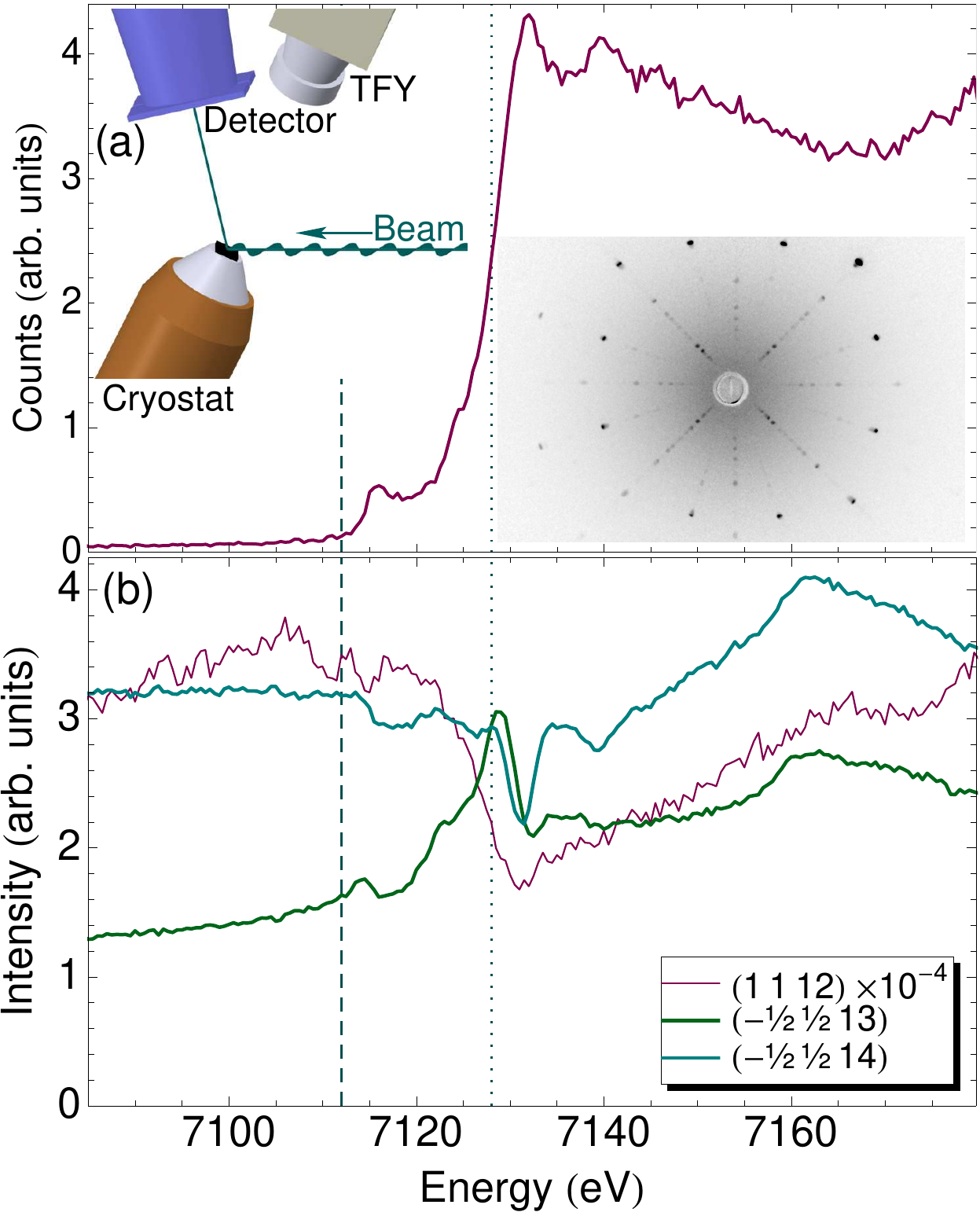}
  \caption{\label{XAS}(a) X-ray absorption spectrum of \SFO\ at the Fe K edge, from total fluorescence yield measurements. The dashed line indicates the beam energy of 7112\,eV used for the temperature-dependent data in the main text while the dotted line identifies the on-resonant energy whose $L$-dependence is fit in Fig. 4(b). The left inset shows the experimental geometry for the ($\frac12$\,$\frac12$\,14) superstructure peak and for fluorescence measurements, while the right inset shows an x-ray Laue pattern of this crystal, viewed along (0\,0\,$L$).  The incoming beam is polarized horizontally (perpendicular to the scattering plane). (b) Energy scans across the Fe K edge of two superstructure peaks and a regular Bragg peak, indexed in $I4/mmm$.  The (1\,1\,12)$_{I4/mmm}$ Bragg peak behaves as expected, with a significant drop in intensity at the resonance followed by a gradual recovery which is due in part to fluorescence.  The superstructure peaks exhibit distinct and nontrivial energy dependence.}
\end{figure}

To enhance the contrast between the two Fe sites, we performed temperature-dependent single-crystal resonant elastic x-ray scattering (REXS) at the Fe $K$ edge at beamline P09 at PETRA III (DESY, in Hamburg, Germany)~\cite{Strempfer2013}, with the experimental geometry shown in the upper inset to Fig.~\ref{XAS}(a). Before measurement, the sample was aligned using a Photonic Science Laue diffractometer with a tungsten source --- the sharp diffraction spots, seen in the inset to Fig.~\ref{XAS}(a), indicate the high quality of the single crystal.  The Laue pattern produced no clear evidence for superstructure reflections. The sample was mounted on the cold finger of a closed-cycle displex cryostat sitting in a six-circle diffractometer, with the (0\,0\,1) axis in the vertical scattering plane; $\sigma$\ polarization was used.  An avalanche photodiode point detector was used to measure the scattered x-ray intensity, and a {\sc Vortex} Si-drift diode fluorescence detector was employed to measure the total fluorescence yield (TFY) from the sample. Fig.~\ref{XAS}(a) shows the TFY of \SFO\ measured as a function of the incident photon energy across the Fe K edge. The main edge at around 7128\,eV and a small pre-edge at around 7115\,eV are clearly visible.

\begin{table*}[htb]
  \caption{\label{BigTable}List of positions probed in reciprocal space referenced to the high-temperature $I4/mmm$ unit cell, together with the peak intensity found there, normalized by incoming beam intensity. Weak peaks were also observed at positions which correspond to intergrowths of the single-layer Ruddlesden-Popper material Sr$_2$FeO$_4$ (which are common in Ruddlesden-Popper phases). From the peak intensities, we estimate that these intergrowths occur below one part in $10^4$. These peaks were verified to retain their full intensity at 330-350\,K, well above the charge order transition in Sr$_3$Fe$_2$O$_7$. Based on the $l$-coordinate and temperature dependence, they can be discriminated from superlattice reflections that belong to Sr$_3$Fe$_2$O$_7$. However, some overlap with Sr$_3$Fe$_2$O$_7$ reflections is noted in the table.  Since this experiment was intended to survey which peaks were present and was not intended to produce quantitative data, readers are cautioned against treating these numbers as quantitative.  In particular, a variety of energies and temperatures were used, and some data were collected using an area detector with different sensitivity, as noted in the table.  The small beam spot will lead to substantial variations in sample illumination and absorption as a function of the angle between the sample's surface normal and the beam, and, even on a single peak, variations at the 10\,\%\ level routinely arise. Peak intensities measured under different conditions should not be compared, and scans along different axes should be treated with caution.}
  \begin{tabular}{l r r c r l}\hline\hline
    Position & \multicolumn{1}{c}{Peak} & \multicolumn{1}{c}{$T$} & Axis & Energy & Notes\\
    & (counts/s) & (K) & & \multicolumn{1}{c}{(eV)} & \\ \hline
    (0\,0\,4) & Present & 5 & $z$ & 7112 & Area detector\\
    (0\,0\,6) & $9.5\times10^{10}$ & 5 & $\theta$ & 7012 &\\
    $\left(\frac12\,\frac12\,6\right)$ & $6.6\times10^5$ & 5 & $L$ & 7112 &\\
    $\left(\frac12\,\frac12\,7\right)$ & $7.4\times10^5$ & 5 & $L$ & 7112 &\\
    (1\,0\,7) & $1.8\times10^8$ & 296& $\theta$ & 7112 & Area detector\\
    (1\,1\,7) & $<600$ & 6 & $\chi$ & 7012 &\\
    $\left(\frac12\,\frac12\,8\right)$ & $6.4\times10^5$ & 5 & $L$ & 7112 &\\
    (1\,1\,8) & $3.9\times10^8$ & 6 & $\chi$ & 7112 & Area detector\\
    $\left(\frac12\,\frac12\,9\right)$ & $9.2\times10^5$ & 5 & $L$ & 7112 &\\
    $\left(\frac12\,\frac32\,9\right)$ & $1.8\times10^5$ & 5 & $L$ & 7112 &\\
    (1\,1\,9) & $<500$ & 6 & $\chi$ & 7012 & \\
    (0\,0\,10) & $1.4\times10^{11}$ & 5 & $\chi$ & 7012 &\\
    $\left(\frac12\,\frac12\,10\right)$ & $1.8\times10^6$ & 5 & $L$ & 7112 &\\
    $\left(\frac12\,\frac32\,10\right)$ & $8.5\times10^5$ & 5 & $L$ & 7112 &\\
    (1\,1\,10) & $2.1\times10^{9}$ & 296 & $\chi$ & 7112 &\\
    $\left(-\frac12\,\frac12\,11\right)$ & $1.2\times10^5$ & 5 & $L$ & 7117 &\\
    $\left(\frac12\,\frac12\,11\right)$ & $6.6\times10^5$ & 25 & $L$ & 7112 &\\
    $\left(\frac12\,\frac32\,11\right)$ & $3.5\times10^5$ & 5 & $L$ & 7112 &\\
    (1\,1\,11) & $<1.5\times10^5$ & 196 & $L$ & 7112 & overlap Sr$_2$FeO$_4$ (1\,1\,7)\\ 
    $\left(-\frac12\,\frac12\,12\right)$ & $8.3\times10^4$ & 5 & $L$ & 7117 &\\
    (0\,0\,12) & $2.1\times10^{10}$ & 5 & $\chi$ & 7012 &\\
    (0\,1\,12) & Absent & 5 & $L$ & 7112 &\\
    $\left(\frac12\,\frac12\,12\right)$ & $6.2\times10^3$ & 5 & $\chi$ & 7012 &\\
    $\left(\frac12\,\frac32\,12\right)$ & $9.5\times10^4$ & 5 & $L$ & 7112 &\\
 %   (1\,0\,12) & $8.2\times10^4$ & 5 & $\chi$ & 7112 & Area detector\\
    (1\,1\,12) & $2.9\times10^{10}$ & 5 & $\chi$ & 7012 &\\
    $\left(1\,1\,12\frac12\right)$ & Absent & 5 & $L$ & 7112 &\\
    $\left(-\frac12\,\frac12\,13\right)$ & $8.4\times10^5$ & 5 & $\chi$ & 7112 &\\
  \hline\hline\end{tabular}~~~~
  \begin{tabular}{l r r c r l}\hline\hline
    Position & \multicolumn{1}{c}{Peak} & \multicolumn{1}{c}{$T$} & Axis & Energy & Notes\\
    & (counts/s) & (K) & & \multicolumn{1}{c}{(eV)} & \\ \hline
    (0\,0\,13) & $< 8.3\times10^5$ & 5 & $\chi$ & 7112 & overlap Sr$_2$FeO$_4$ (0\,0\,8)\\
    (0\,1\,13) & $3.5\times10^9$ & 5 & $\chi$ & 7112 &\\
    $\left(\frac12\,\frac12\,13\right)$ & $7.8\times10^5$ & 5 & $\chi$ & 7112 &\\
    $\left(\frac12\,\frac32\,13\right)$ & $2.0\times10^5$ & 5 & $\chi$ & 7112 &\\
    (1\,0\,13) & $4.3\times10^8$ & 5 & $\chi$ & 7132 &\\
    (1\,1\,13) & $<2.0\times10^4$ & 196 & $L$ & 7112 & overlap Sr$_2$FeO$_4$ (1\,1\,8)\\ 
    $\left(0\,0\,13\frac12\right)$ & Absent & 5 & $L$ & 7112 &\\
    $\left(1\,1\,13\frac12\right)$ & Absent & 5 & $L$ & 7112 &\\
    $\left(-\frac12\,\frac12\,14\right)$ & $1.8\times10^6$ & 5 & $L$ & 7117 &\\
    (0\,1\,14) & Absent & 5 & $L$ & 7112 &\\
    $\left(\frac12\,\frac12\,14\right)$ & $2.3\times10^6$ & 25 & $\chi$ & 7112 &\\
    $\left(\frac12\,\frac32\,14\right)$ & $9.0\times10^5$ & 6 & $\chi$ & 7112 &\\
    (1\,1\,14) & $2.2\times10^{10}$ & 5 & $L$ & 7112 &\\
    $\left(1\,1\,14\frac12\right)$ & Absent & 5 & $L$ & 7112 &\\
    $\left(-\frac12\,\frac12\,15\right)$ & $8.6\times10^5$ & 5 & $\chi$ & 7112 &\\
    (0\,1\,15) & $5.4\times10^{10}$ & 5 & $\chi$ & 7112 &\\
    $\left(\frac12\,\frac12\,15\right)$ & $1.0\times10^6$ & 25 & $L$ & 7112 &\\
    $\left(\frac12\,\frac32\,15\right)$ & $4.1\times10^5$ & 5 & $L$ & 7112 &\\
    $\left(0\,1\,15\frac12\right)$ & Absent & 5 & $L$ & 7112 &\\
    $\left(-\frac12\,\frac12\,16\right)$ & $1.9\times10^5$ & 5 & $L$ & 7080 &\\
    (0\,1\,16) & $< 1.0\times10^6$ & 196 & $L$ & 7112 & overlap Sr$_2$FeO$_4$ (0\,1\,10)\\
    $\left(\frac12\,\frac12\,16\right)$ & $6.2\times10^5$ & 25 & $L$ & 7112 &\\
    $\left(\frac12\,\frac32\,16\right)$ & $2.9\times10^5$ & 5 & $L$ & 7112 &\\
    $\left(0\,1\,16\frac12\right)$ & Absent & 197 & $L$ & 7112 &\\
    $\left(-\frac12\,\frac12\,17\right)$ & $5.4\times10^5$ & 5 & $L$ & 7117 &\\    
    $\left(\frac12\,\frac12\,17\right)$ & $9.1\times10^5$ & 25 & $L$ & 7112 &\\
    ~ & & & & & \\
  \hline\hline\end{tabular}\hfill
\end{table*}

As can be seen in Fig.~\ref{XAS}(b), which references peaks to the $I4/mmm$ cell, the diffraction anomalous fine structure (DAFS) of the superstructure peaks is complex. While the regular tetragonal Bragg peak (1\,1\,12) shows a characteristic absorption dip at the edge, the two superstructure peaks exhibit a complex energy-dependence throughout the entire edge region. This is related to interference effects between the two Fe sites with  different valences in combination with the short correlation length along the $L$ direction.  The temperature-dependent REXS data presented in the main text were collected at an energy of 7112\,eV, or 1.743\,\AA, corresponding to the dashed vertical line in Fig.~\ref{XAS} just below the pre-edge features. This energy avoids the strong fluorescence background but is close enough to the edge in order to benefit from resonant enhancement.  Table \ref{BigTable} lists reciprocal-space positions probed by resonant x-ray scattering.

 \section{M\"ossbauer spectroscopy}

M\"ossbauer spectra were collected between 4.8 and 316 K with a standard WissEl spectrometer, which was operated in constant-acceleration mode and was equipped with a $^{57}$Co/Rh source. For the absorber, a crystal of Sr$_3$Fe$_2$O$_7$ was ground. The powder containing about 10\,mg of natural Fe/cm$^2$ was diluted with boron nitride to ensure homogeneous distribution and filled into a Plexiglas sample container. In order to prevent sample degradation by moisture the absorber was prepared in an Ar-filled glovebox. Spectra at different temperatures were obtained using a Janis SHI-850-5 closed cycle refrigerator (CCR); the spectra at 292 and 316\,K were collected with the CCR switched off. Isomer shifts are given relative to $\alpha$-Fe. The data were evaluated with the program {\sc MossWinn} \cite{Mosswinn} in the perturbation limit $QS\ll B_{hf}$, where $QS$ corresponds to the quadrupole splitting and $B_{hf}$ to the hyperfine field. Spectra were evaluated with Lorentzian-type sextets or with hyperfine field distributions according to the Hesse-R\"ubartsch method.

\begin{figure}[htb]
  \includegraphics[width=\columnwidth]{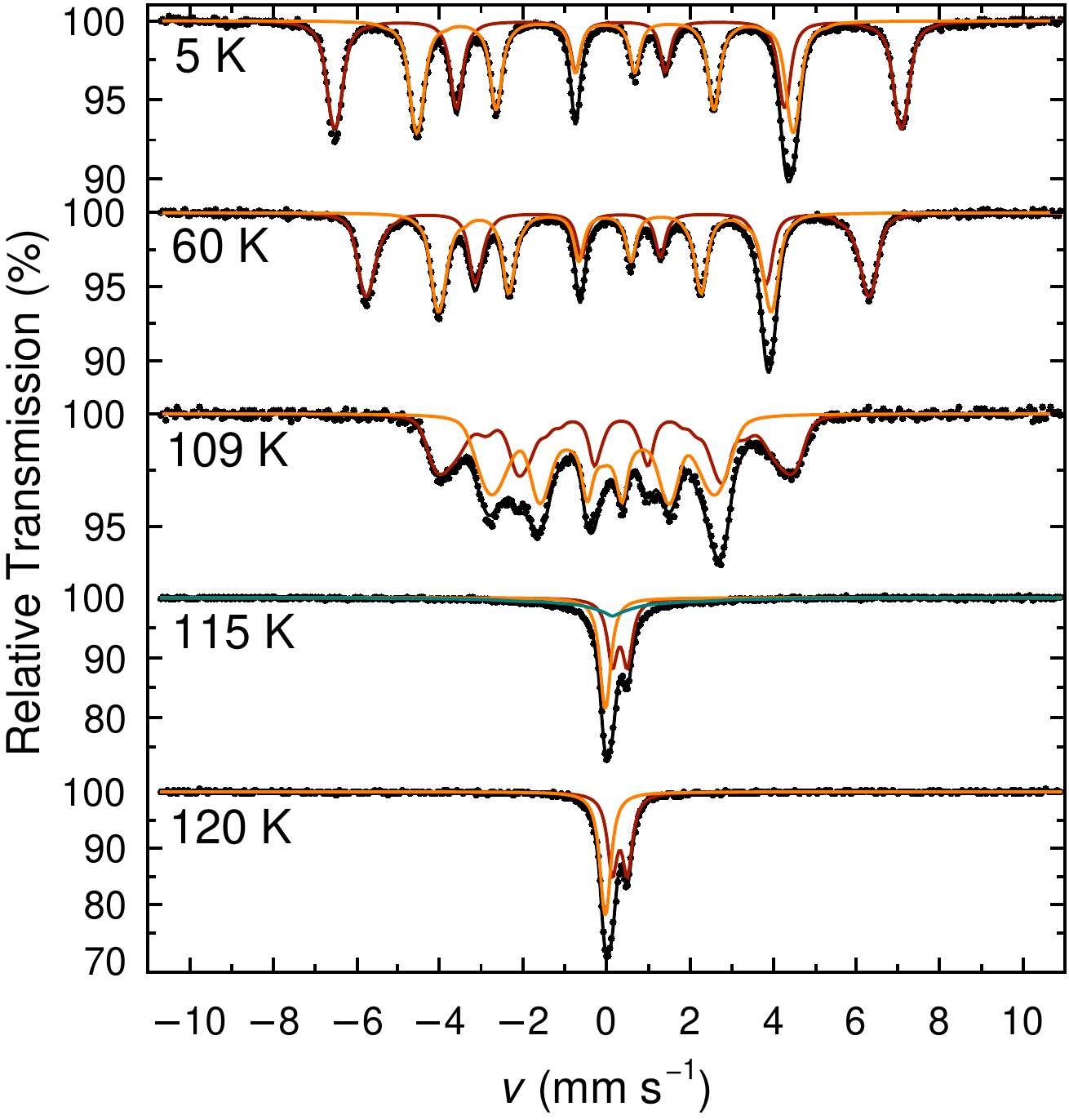}
  \caption{\label{Moess2}M\"ossbauer spectra of \SFO\ in the magnetically-ordered phase and around the magnetic phase transition. The dark and light components correspond to the ``Fe$^{3+}$'' and ``Fe$^{5+}$'' sites, respectively. The blue component in the 115-K spectrum reflects the residual magnetically-ordered phase.}
\end{figure}

\begin{figure}[htb]
  \includegraphics[width=\columnwidth]{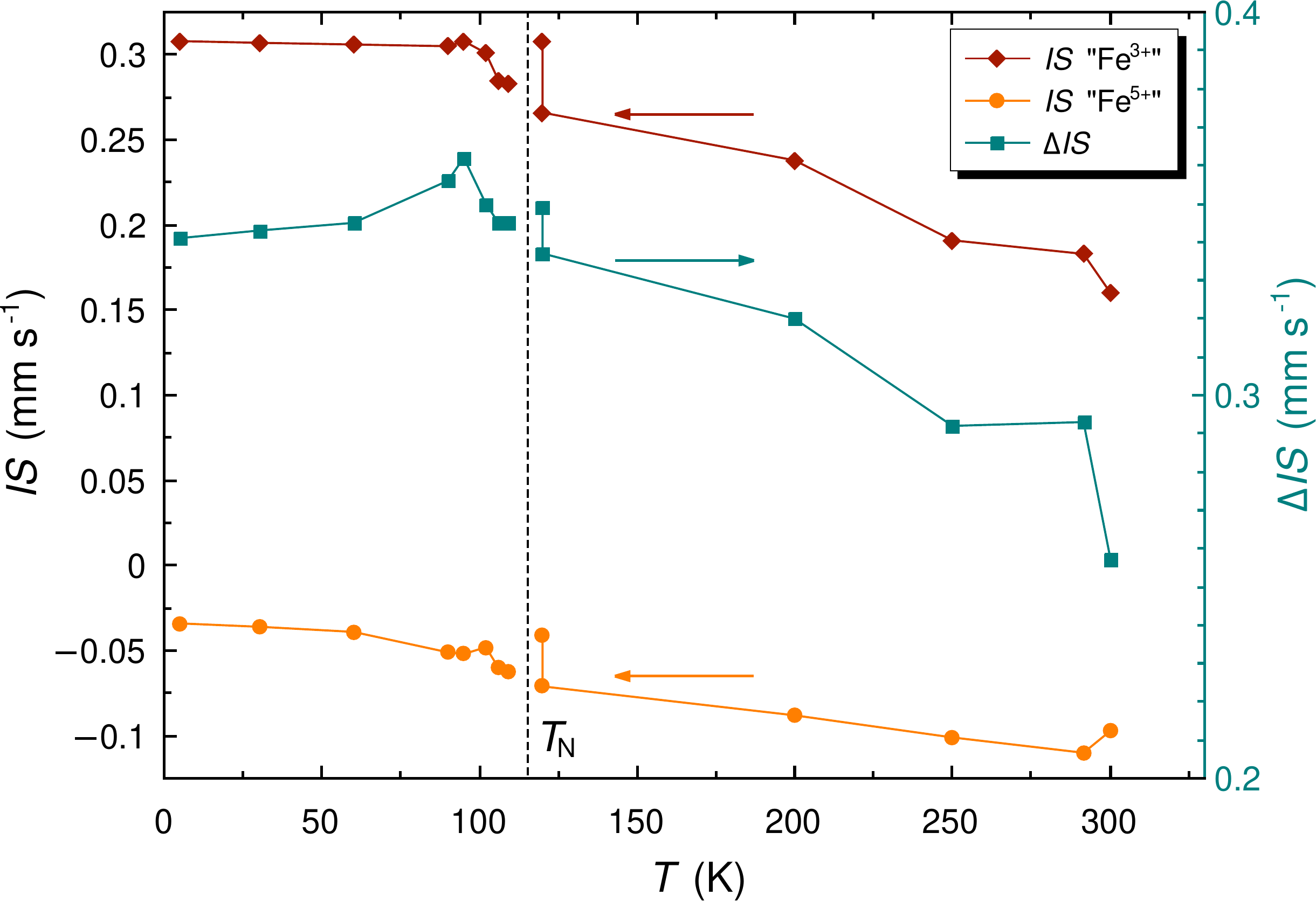}
  \caption{\label{MoessIS}Temperature dependence of the isomer shifts $IS$ of the ``Fe$^{3+}$'' (dark) and ``Fe$^{5+}$'' (light) sites and of the difference $\Delta IS$ of the isomer shifts between the two sites (green). The solid lines are guides to the eye.}
\end{figure}

\begin{figure}[htb]
  \includegraphics[width=\columnwidth]{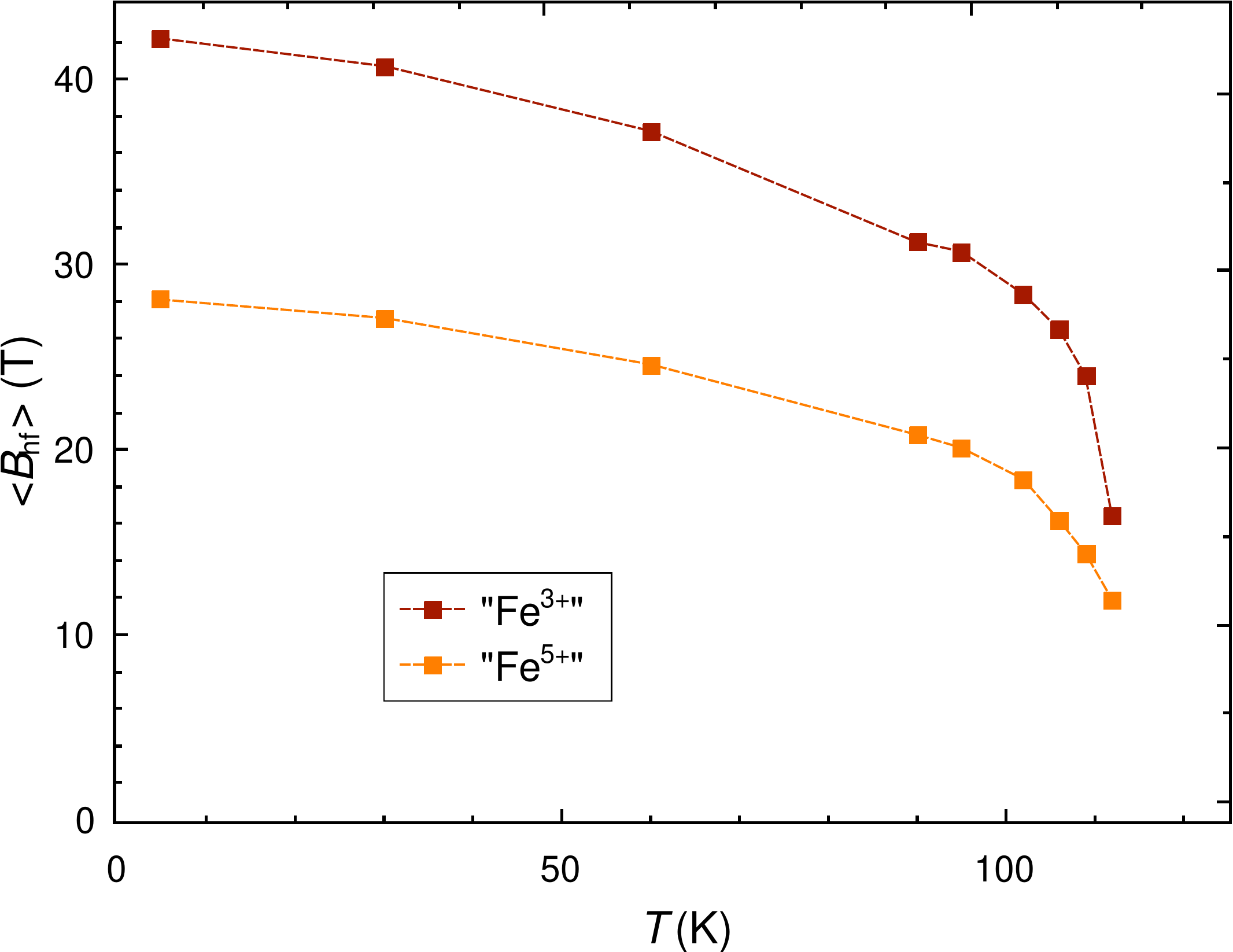}
  \caption{\label{MoessHF}Temperature dependence of the average hyperfine fields $\langle B_{hf} \rangle$ of the ``Fe$^{3+}$'' (dark) and ``Fe$^{5+}$'' (light) sites. The data were obtained by evaluation of the spectra with a distribution model. The N\'eel temperature was chosen to be 115.0\,K, although a minor fraction is still in the magnetically ordered phase. The dashed lines are guides to the eye.}
\end{figure}

Representative M\"ossbauer spectra of \SFO\ in the temperature range 5-120\,K are shown in Fig.~\ref{Moess2}, and the isomer shifts $IS$ and hyperfine fields $B_{hf}$ are depicted in Figs.~\ref{MoessIS} and \ref{MoessHF}. The spectrum at 5\,K can be described by two hyperfine sextets with distinct $IS$ and $B_{hf}$ values which correspond to two inequivalent iron sites. The two sextets are somewhat broadened.  Since the line broadening increases with temperature, the spectra were described by two distributions of hyperfine fields, rather than by two distinct sextets with Lorentzian lineshape. The area ratio between the two subspectra is 1:1 which confirms a charge disproportionation (CD) of Fe$^{4+}$, in agreement with previous investigations of highly-oxidized \SFO\cite{Dann1993,Adler1997,Kuzushita2000}.

The CD is frequently written as 2Fe$^{4+}\rightarrow$ Fe$^{3+}$ + Fe$^{5+}$. The $IS$ and $B_{hf}$ values are similar to those in CaFeO$_3$ \cite{Takano1977} which is the classical example for a CD of Fe$^{4+}$. However, as pointed out previously\cite{Adler1997}, the $IS$ of 0.31\,mm/s at 5\,K for the ``Fe$^{3+}$'' site is considerably smaller than the typical values of $\sim$0.45\,mm/s for Fe$^{3+}$ in octahedral oxygen coordination, whereas the $IS$ = $-$0.03\,mm/s for ``Fe$^{5+}$'' is larger than for instance $IS$ = $-$0.34\,mm/s found in the Fe$^{5+}$ double perovskite La$_2$FeLiO$_6$\cite{Demazeau1981}. Similarly, the hyperfine fields of 42 and 28\,T are smaller and larger than expected values of $>$50\,T for Fe$^{3+}$ and 23\,T for Fe$^{5+}$, respectively\cite{Demazeau1981}. Accordingly, the differences $\Delta IS$ and $\Delta B_{hf}$ in isomer shifts and hyperfine fields between the two species are smaller than expected for a full charge disproportionation, which is in qualitative agreement with the view that these formal iron(IV) oxides are strongly covalent and can be considered as negative-$\Delta$ materials\cite{Bocquet1992}, where $\Delta$ is the charge-transfer energy. Then the CD may be formulated as 2\,$d^5$L$^{-1}$ $\rightarrow$ $d^5$ + $d^5$L$^{-2}$, where L$^{-1}$ and L$^{-2}$ represent one and two holes in the oxygen coordination sphere, respectively. The formulation 2Fe$^{4+}$ $\rightarrow$ Fe$^{(4-\delta)+}$ + Fe$^{(4+\delta)+}$ has also been used to indicate an incomplete degree of charge separation. In any case, the CD clearly alters the electron and spin densities at the iron sites which gives rise to distinct $IS$ and $B_{hf}$ values. As there is only a small quadrupole interaction, the different $B_{hf}$s correspond to different ordered magnetic moments at the two sites. For CaFeO$_3$, ordered magnetic moments of 3.5 and 2.5\,$\mu_B$ were obtained from neutron diffraction studies for the helical state of CaFeO$_3$\cite{Woodward2000}. Since the $B_{hf}$ values of CaFeO$_3$ are quite similar to those of \SFO, similar moments are expected. The difference $\Delta IS$ is also nearly the same in the two compounds, which implies a comparable degree of charge segregation in \SFO\ and CaFeO$_3$.

The CD remains nearly unchanged up to the magnetic ordering temperature $T_\text{N}$ $\sim$115\,K, where phase coexistence of the paramagnetic and magnetically-ordered phases is found. The latter appears as a broad magnetic hyperfine pattern which is superimposed on the paramagnetic subspectrum. In the magnetically-ordered phase the spectra become continuously broadened with increasing temperature which is reflected in an increased distribution width of $B_{hf}$. The broadening could reflect slight variations of the magnetic ordering temperature due to a residual oxygen deficiency and/or spin fluctuations. It is noteworthy that even at 5\,K a broadening is observed which is more pronounced for the ``Fe$^{3+}$'' subspectrum than for the ``Fe$^{5+}$'' subspectrum. This may reflect the helical spin structure of \SFO\cite{Kim2014}. The ``Fe$^{3+}$'' site has a larger quadrupole splitting than the ``Fe$^{5+}$'' site, where the quadrupole interaction essentially vanishes. The quadrupole splitting parameter $QS$ for ``Fe$^{3+}$'' ($-$0.06\,mm/s at 5\,K) depends on the angle between the principal axis $V_{ZZ}$ of the electric field gradient (efg) and the spin direction, which varies in the case of a helical spin structure. The resulting distribution in $QS$ may be the origin of the increased broadening in the case of the ``Fe$^{3+}$'' component. 

\begin{figure}[htb]
  \includegraphics[width=\columnwidth]{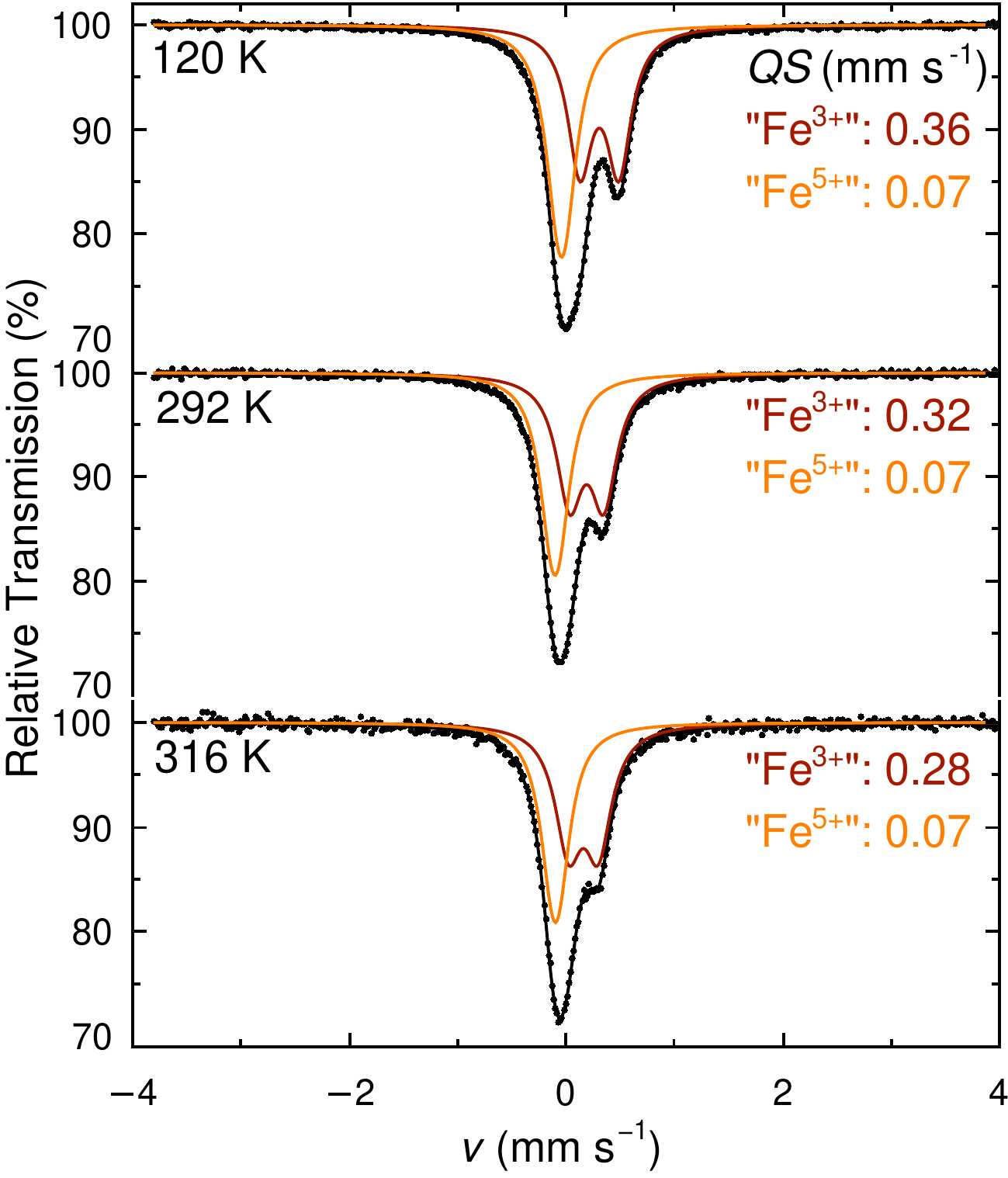}
  \caption{\label{MoessHigh}M\"ossbauer spectra of \SFO\ in the paramagnetic phase. The dark and light components correspond to the ``Fe$^{3+}$'' and ``Fe$^{5+}$'' sites, respectively. The values $QS$ of the quadrupole splitting are given in the figure.}
\end{figure}

At 120\,K the sample is completely in the paramagnetic state. It is obvious that the spectra between 120 and 316\,K (Fig.~\ref{MoessHigh}) are composed of two components, which evidences that the CD persists in the paramagnetic phase. The spectra were described by two quadrupole doublets, with quadrupole splitting $QS$ of 0.07 and 0.36\,mm/s at 120\,K for the ``Fe$^{5+}$'' and ``Fe$^{3+}$'' sites, respectively. Thus, the ``Fe$^{5+}$'' component is essentially a slightly broadened single line. The temperature dependence of the isomer shifts and of the difference between the isomer shifts $\Delta IS$ of the two sites in both the magnetically ordered and the paramagnetic phase is shown in Fig.~\ref{MoessIS}. In agreement with the earlier results these data suggest that the CD is largely insensitive to the magnetic phase transition. With increasing temperature, in particular above 200\,K, $\Delta IS$ decreases, which indicates that the degree of charge separation is reduced\cite{Adler1997}, an effect which is also visible in the resonant diffraction results in the main text. Nevertheless, the CD is still apparent at 316\,K, the highest temperature in this study. This is in agreement with the results in the main text and with the slightly higher $T_\text{CO}$ = 343$\pm$10\,K reported in Ref.~\onlinecite{Kuzushita2000}. 

We note that the fit of the spectra in the paramagnetic phase is not unique, and there is another fit with somewhat increased values of $QS$ for both the ``Fe$^{3+}$'' and the ``Fe$^{5+}$'' sites which reproduces the spectra equally well. This data analysis, however, results in an anomaly in $\Delta IS$ near the magnetic phase transition and thus would point to a magnetostriction effect, which reduces the charge segregation at $T_\text{N}$. A magnetostriction effect is, however, excluded by the neutron Larmor diffraction results in the main text, and thus this fit is discarded.

In summary, the M\"ossbauer study on the present \SFO\ sample corroborates the CD of Fe$^{4+}$ and is fully consistent with the crystal structure data at 15\,K which reveal a bond-length alternation within in the double layers. The Fe1 sites with the larger bond distances correspond to the ``Fe$^{3+}$'' sites in the M\"ossbauer spectra, whereas the Fe2 sites correspond to the contracted ``Fe$^{5+}$” sites.

%% file: Sr3Fe2O7_final.bbl
%apsrev4-2.bst 2019-01-14 (MD) hand-edited version of apsrev4-1.bst
%Control: key (0)
%Control: author (8) initials jnrlst
%Control: editor formatted (1) identically to author
%Control: production of article title (0) allowed
%Control: page (0) single
%Control: year (1) truncated
%Control: production of eprint (0) enabled
\begin{thebibliography}{46}%
\makeatletter
\providecommand \@ifxundefined [1]{%
 \@ifx{#1\undefined}
}%
\providecommand \@ifnum [1]{%
 \ifnum #1\expandafter \@firstoftwo
 \else \expandafter \@secondoftwo
 \fi
}%
\providecommand \@ifx [1]{%
 \ifx #1\expandafter \@firstoftwo
 \else \expandafter \@secondoftwo
 \fi
}%
\providecommand \natexlab [1]{#1}%
\providecommand \enquote  [1]{``#1''}%
\providecommand \bibnamefont  [1]{#1}%
\providecommand \bibfnamefont [1]{#1}%
\providecommand \citenamefont [1]{#1}%
\providecommand \href@noop [0]{\@secondoftwo}%
\providecommand \href [0]{\begingroup \@sanitize@url \@href}%
\providecommand \@href[1]{\@@startlink{#1}\@@href}%
\providecommand \@@href[1]{\endgroup#1\@@endlink}%
\providecommand \@sanitize@url [0]{\catcode `\\12\catcode `\$12\catcode
  `\&12\catcode `\#12\catcode `\^12\catcode `\_12\catcode `\%12\relax}%
\providecommand \@@startlink[1]{}%
\providecommand \@@endlink[0]{}%
\providecommand \url  [0]{\begingroup\@sanitize@url \@url }%
\providecommand \@url [1]{\endgroup\@href {#1}{\urlprefix }}%
\providecommand \urlprefix  [0]{URL }%
\providecommand \Eprint [0]{\href }%
\providecommand \doibase [0]{https://doi.org/}%
\providecommand \selectlanguage [0]{\@gobble}%
\providecommand \bibinfo  [0]{\@secondoftwo}%
\providecommand \bibfield  [0]{\@secondoftwo}%
\providecommand \translation [1]{[#1]}%
\providecommand \BibitemOpen [0]{}%
\providecommand \bibitemStop [0]{}%
\providecommand \bibitemNoStop [0]{.\EOS\space}%
\providecommand \EOS [0]{\spacefactor3000\relax}%
\providecommand \BibitemShut  [1]{\csname bibitem#1\endcsname}%
\let\auto@bib@innerbib\@empty
%</preamble>
\bibitem [{\citenamefont {Mydosh}\ and\ \citenamefont
  {Oppeneer}(2011)}]{Mydosh2011}%
  \BibitemOpen
  \bibfield  {author} {\bibinfo {author} {\bibfnamefont {J.~A.}\ \bibnamefont
  {Mydosh}}\ and\ \bibinfo {author} {\bibfnamefont {P.~M.}\ \bibnamefont
  {Oppeneer}},\ }\bibfield  {title} {\bibinfo {title} {Colloquium: Hidden
  order, superconductivity, and magnetism: The unsolved case of
  {URu}$_2${Si}$_2$},\ }\href {https://doi.org/10.1103/RevModPhys.83.1301}
  {\bibfield  {journal} {\bibinfo  {journal} {Rev. Mod. Phys.}\ }\textbf
  {\bibinfo {volume} {83}},\ \bibinfo {pages} {1301} (\bibinfo {year}
  {2011})},\ \Eprint {https://arxiv.org/abs/1107.0258} {arXiv:1107.0258
  [cond-mat.supr-con]} \BibitemShut {NoStop}%
\bibitem [{\citenamefont {Mydosh}\ \emph {et~al.}(2020)\citenamefont {Mydosh},
  \citenamefont {Oppeneer},\ and\ \citenamefont {Riseborough}}]{Mydosh2020}%
  \BibitemOpen
  \bibfield  {author} {\bibinfo {author} {\bibfnamefont {J.~A.}\ \bibnamefont
  {Mydosh}}, \bibinfo {author} {\bibfnamefont {P.~M.}\ \bibnamefont
  {Oppeneer}},\ and\ \bibinfo {author} {\bibfnamefont {P.~S.}\ \bibnamefont
  {Riseborough}},\ }\bibfield  {title} {\bibinfo {title} {Hidden order and
  beyond: an experimental---theoretical overview of the multifaceted behavior
  of {URu}$_2${Si}$_2$},\ }\href {https://doi.org/10.1088/1361-648x/ab5eba}
  {\bibfield  {journal} {\bibinfo  {journal} {Journal of Physics: Condensed
  Matter}\ }\textbf {\bibinfo {volume} {32}},\ \bibinfo {pages} {143002}
  (\bibinfo {year} {2020})},\ \Eprint {https://arxiv.org/abs/1912.09145}
  {arXiv:1912.09145 [cond-mat.str-el]} \BibitemShut {NoStop}%
\bibitem [{\citenamefont {Cameron}\ \emph {et~al.}(2016)\citenamefont
  {Cameron}, \citenamefont {Friemel},\ and\ \citenamefont
  {Inosov}}]{Cameron2016}%
  \BibitemOpen
  \bibfield  {author} {\bibinfo {author} {\bibfnamefont {A.~S.}\ \bibnamefont
  {Cameron}}, \bibinfo {author} {\bibfnamefont {G.}~\bibnamefont {Friemel}},\
  and\ \bibinfo {author} {\bibfnamefont {D.~S.}\ \bibnamefont {Inosov}},\
  }\bibfield  {title} {\bibinfo {title} {Multipolar phases and magnetically
  hidden order: review of the heavy-fermion compound
  {Ce}$_{1-x}${La}$_x${B}$_6$},\ }\href
  {https://doi.org/10.1088/0034-4885/79/6/066502} {\bibfield  {journal}
  {\bibinfo  {journal} {Rep.\ Prog.\ Phys.}\ }\textbf {\bibinfo {volume}
  {79}},\ \bibinfo {pages} {066502} (\bibinfo {year} {2016})},\ \Eprint
  {https://arxiv.org/abs/1509.03588} {arXiv:1509.03588 [cond-mat.str-el]}
  \BibitemShut {NoStop}%
\bibitem [{\citenamefont {Cao}\ and\ \citenamefont
  {Schlottmann}(2018)}]{Cao2018}%
  \BibitemOpen
  \bibfield  {author} {\bibinfo {author} {\bibfnamefont {G.}~\bibnamefont
  {Cao}}\ and\ \bibinfo {author} {\bibfnamefont {P.}~\bibnamefont
  {Schlottmann}},\ }\bibfield  {title} {\bibinfo {title} {The challenge of
  spin--orbit-tuned ground states in iridates: a key issues review},\ }\href
  {https://doi.org/10.1088/1361-6633/aaa979} {\bibfield  {journal} {\bibinfo
  {journal} {Rep. Prog. Phys.}\ }\textbf {\bibinfo {volume} {81}},\ \bibinfo
  {pages} {042502} (\bibinfo {year} {2018})},\ \Eprint
  {https://arxiv.org/abs/1704.06007} {arXiv:1704.06007 [cond-mat.str-el]}
  \BibitemShut {NoStop}%
\bibitem [{\citenamefont {Gallagher}\ \emph {et~al.}(1966)\citenamefont
  {Gallagher}, \citenamefont {MacChesney},\ and\ \citenamefont
  {Buchanan}}]{Gallagher1966}%
  \BibitemOpen
  \bibfield  {author} {\bibinfo {author} {\bibfnamefont {P.~K.}\ \bibnamefont
  {Gallagher}}, \bibinfo {author} {\bibfnamefont {J.~B.}\ \bibnamefont
  {MacChesney}},\ and\ \bibinfo {author} {\bibfnamefont {D.~N.~E.}\
  \bibnamefont {Buchanan}},\ }\bibfield  {title} {\bibinfo {title}
  {M{\"o}ssbauer effect in the system {Sr$_3$Fe$_2$O$_{6-7}$}},\ }\href
  {https://doi.org/10.1063/1.1727962} {\bibfield  {journal} {\bibinfo
  {journal} {J. Chem. Phys.}\ }\textbf {\bibinfo {volume} {45}},\ \bibinfo
  {pages} {2466} (\bibinfo {year} {1966})}\BibitemShut {NoStop}%
\bibitem [{\citenamefont {Dann}\ \emph {et~al.}(1993)\citenamefont {Dann},
  \citenamefont {Weller}, \citenamefont {Currie}, \citenamefont {Thomas},\ and\
  \citenamefont {Al-Rawwas}}]{Dann1993}%
  \BibitemOpen
  \bibfield  {author} {\bibinfo {author} {\bibfnamefont {S.~E.}\ \bibnamefont
  {Dann}}, \bibinfo {author} {\bibfnamefont {M.~T.}\ \bibnamefont {Weller}},
  \bibinfo {author} {\bibfnamefont {D.~B.}\ \bibnamefont {Currie}}, \bibinfo
  {author} {\bibfnamefont {M.~F.}\ \bibnamefont {Thomas}},\ and\ \bibinfo
  {author} {\bibfnamefont {A.~D.}\ \bibnamefont {Al-Rawwas}},\ }\bibfield
  {title} {\bibinfo {title} {Structure and magnetic properties of
  {Sr$_2$FeO$_4$} and {Sr$_3$Fe$_2$O$_7$} studied by powder neutron diffraction
  and {M}{\"o}ssbauer spectroscopy},\ }\href
  {https://doi.org/10.1039/jm9930301231} {\bibfield  {journal} {\bibinfo
  {journal} {J. Mater. Chem.}\ }\textbf {\bibinfo {volume} {3}},\ \bibinfo
  {pages} {1231} (\bibinfo {year} {1993})}\BibitemShut {NoStop}%
\bibitem [{\citenamefont {Adler}(1997)}]{Adler1997}%
  \BibitemOpen
  \bibfield  {author} {\bibinfo {author} {\bibfnamefont {P.}~\bibnamefont
  {Adler}},\ }\bibfield  {title} {\bibinfo {title} {Electronic state,
  magnetism, and electrical transport behavior of {Sr$_{3-x}A_x$Fe$_2$O$_7$}
  ($x \leq 0.4$, {$A$ = Ba, La})},\ }\href
  {https://doi.org/10.1006/jssc.1997.7289} {\bibfield  {journal} {\bibinfo
  {journal} {Journal of Solid State Chemistry}\ }\textbf {\bibinfo {volume}
  {130}},\ \bibinfo {pages} {129} (\bibinfo {year} {1997})}\BibitemShut
  {NoStop}%
\bibitem [{\citenamefont {Kobayashi}\ \emph {et~al.}(1997)\citenamefont
  {Kobayashi}, \citenamefont {Kira}, \citenamefont {Onodera}, \citenamefont
  {Suzuki},\ and\ \citenamefont {Kamimura}}]{Kobayashi1997}%
  \BibitemOpen
  \bibfield  {author} {\bibinfo {author} {\bibfnamefont {H.}~\bibnamefont
  {Kobayashi}}, \bibinfo {author} {\bibfnamefont {M.}~\bibnamefont {Kira}},
  \bibinfo {author} {\bibfnamefont {H.}~\bibnamefont {Onodera}}, \bibinfo
  {author} {\bibfnamefont {T.}~\bibnamefont {Suzuki}},\ and\ \bibinfo {author}
  {\bibfnamefont {T.}~\bibnamefont {Kamimura}},\ }\bibfield  {title} {\bibinfo
  {title} {Electronic state of {Sr$_3$Fe$_2$O$_{7-y}$} studied by specific heat
  and {M}{\"o}ssbauer spectroscopy},\ }\href
  {https://doi.org/10.1016/S0921-4526(97)00065-3} {\bibfield  {journal}
  {\bibinfo  {journal} {Physica B: Condensed Matter}\ }\textbf {\bibinfo
  {volume} {237-238}},\ \bibinfo {pages} {105} (\bibinfo {year} {1997})},\
  \bibinfo {note} {proceedings of the Yamada Conference XLV, the International
  Conference on the Physics of Transition Metals}\BibitemShut {NoStop}%
\bibitem [{\citenamefont {Adler}\ \emph {et~al.}(1999)\citenamefont {Adler},
  \citenamefont {Schwarz}, \citenamefont {Syassen}, \citenamefont {Rozenberg},
  \citenamefont {Machavariani}, \citenamefont {Milner}, \citenamefont
  {Pasternak},\ and\ \citenamefont {Hanfland}}]{Adler1999}%
  \BibitemOpen
  \bibfield  {author} {\bibinfo {author} {\bibfnamefont {P.}~\bibnamefont
  {Adler}}, \bibinfo {author} {\bibfnamefont {U.}~\bibnamefont {Schwarz}},
  \bibinfo {author} {\bibfnamefont {K.}~\bibnamefont {Syassen}}, \bibinfo
  {author} {\bibfnamefont {G.~K.}\ \bibnamefont {Rozenberg}}, \bibinfo {author}
  {\bibfnamefont {G.~Y.}\ \bibnamefont {Machavariani}}, \bibinfo {author}
  {\bibfnamefont {A.~P.}\ \bibnamefont {Milner}}, \bibinfo {author}
  {\bibfnamefont {M.~P.}\ \bibnamefont {Pasternak}},\ and\ \bibinfo {author}
  {\bibfnamefont {M.}~\bibnamefont {Hanfland}},\ }\bibfield  {title} {\bibinfo
  {title} {Collapse of the charge disproportionation and covalency-driven
  insulator-metal transition in {Sr}$_3${Fe}$_2${O}$_7$ under pressure},\
  }\href {https://doi.org/10.1103/PhysRevB.60.4609} {\bibfield  {journal}
  {\bibinfo  {journal} {Phys. Rev. B}\ }\textbf {\bibinfo {volume} {60}},\
  \bibinfo {pages} {4609} (\bibinfo {year} {1999})}\BibitemShut {NoStop}%
\bibitem [{\citenamefont {Mori}\ \emph {et~al.}(1999)\citenamefont {Mori},
  \citenamefont {Kamiyama}, \citenamefont {Kobayashi}, \citenamefont {Torii},
  \citenamefont {Izumi},\ and\ \citenamefont {Asano}}]{Mori1999}%
  \BibitemOpen
  \bibfield  {author} {\bibinfo {author} {\bibfnamefont {K.}~\bibnamefont
  {Mori}}, \bibinfo {author} {\bibfnamefont {T.}~\bibnamefont {Kamiyama}},
  \bibinfo {author} {\bibfnamefont {H.}~\bibnamefont {Kobayashi}}, \bibinfo
  {author} {\bibfnamefont {S.}~\bibnamefont {Torii}}, \bibinfo {author}
  {\bibfnamefont {F.}~\bibnamefont {Izumi}},\ and\ \bibinfo {author}
  {\bibfnamefont {H.}~\bibnamefont {Asano}},\ }\bibfield  {title} {\bibinfo
  {title} {Crystal structure of {Sr$_3$Fe$_2$O$_{7-\delta}$}},\ }\href
  {https://doi.org/10.1016/S0022-3697(99)00158-4} {\bibfield  {journal}
  {\bibinfo  {journal} {Journal of Physics and Chemistry of Solids}\ }\textbf
  {\bibinfo {volume} {60}},\ \bibinfo {pages} {1443} (\bibinfo {year}
  {1999})}\BibitemShut {NoStop}%
\bibitem [{\citenamefont {Kuzushita}\ \emph {et~al.}(2000)\citenamefont
  {Kuzushita}, \citenamefont {Morimoto}, \citenamefont {Nasu},\ and\
  \citenamefont {Nakamura}}]{Kuzushita2000}%
  \BibitemOpen
  \bibfield  {author} {\bibinfo {author} {\bibfnamefont {K.}~\bibnamefont
  {Kuzushita}}, \bibinfo {author} {\bibfnamefont {S.}~\bibnamefont {Morimoto}},
  \bibinfo {author} {\bibfnamefont {S.}~\bibnamefont {Nasu}},\ and\ \bibinfo
  {author} {\bibfnamefont {S.}~\bibnamefont {Nakamura}},\ }\bibfield  {title}
  {\bibinfo {title} {Charge disproportionation and antiferromagnetic order of
  {Sr$_3$Fe$_2$O$_7$}},\ }\href {https://doi.org/10.1143/JPSJ.69.2767}
  {\bibfield  {journal} {\bibinfo  {journal} {Journal of the Physical Society
  of Japan}\ }\textbf {\bibinfo {volume} {69}},\ \bibinfo {pages} {2767}
  (\bibinfo {year} {2000})}\BibitemShut {NoStop}%
\bibitem [{\citenamefont {Peets}\ \emph {et~al.}(2013)\citenamefont {Peets},
  \citenamefont {Kim}, \citenamefont {Dosanjh}, \citenamefont {Reehuis},
  \citenamefont {Maljuk}, \citenamefont {Aliouane}, \citenamefont {Ulrich},\
  and\ \citenamefont {Keimer}}]{Peets2013}%
  \BibitemOpen
  \bibfield  {author} {\bibinfo {author} {\bibfnamefont {D.~C.}\ \bibnamefont
  {Peets}}, \bibinfo {author} {\bibfnamefont {J.-H.}\ \bibnamefont {Kim}},
  \bibinfo {author} {\bibfnamefont {P.}~\bibnamefont {Dosanjh}}, \bibinfo
  {author} {\bibfnamefont {M.}~\bibnamefont {Reehuis}}, \bibinfo {author}
  {\bibfnamefont {A.}~\bibnamefont {Maljuk}}, \bibinfo {author} {\bibfnamefont
  {N.}~\bibnamefont {Aliouane}}, \bibinfo {author} {\bibfnamefont
  {C.}~\bibnamefont {Ulrich}},\ and\ \bibinfo {author} {\bibfnamefont
  {B.}~\bibnamefont {Keimer}},\ }\bibfield  {title} {\bibinfo {title} {Magnetic
  phase diagram of {Sr}$_3${Fe}$_2${O}$_{7-\delta}$},\ }\href
  {https://doi.org/10.1103/PhysRevB.87.214410} {\bibfield  {journal} {\bibinfo
  {journal} {Phys. Rev. B}\ }\textbf {\bibinfo {volume} {87}},\ \bibinfo
  {pages} {214410} (\bibinfo {year} {2013})},\ \Eprint
  {https://arxiv.org/abs/1302.1815} {arXiv:1302.1815 [cond-mat.str-el]}
  \BibitemShut {NoStop}%
\bibitem [{\citenamefont {Rekveldt}\ \emph {et~al.}(2001)\citenamefont
  {Rekveldt}, \citenamefont {Keller},\ and\ \citenamefont
  {Golub}}]{Rekveldt2001}%
  \BibitemOpen
  \bibfield  {author} {\bibinfo {author} {\bibfnamefont {M.~T.}\ \bibnamefont
  {Rekveldt}}, \bibinfo {author} {\bibfnamefont {T.}~\bibnamefont {Keller}},\
  and\ \bibinfo {author} {\bibfnamefont {R.}~\bibnamefont {Golub}},\ }\bibfield
   {title} {\bibinfo {title} {Larmor precession, a technique for
  high-sensitivity neutron diffraction},\ }\href
  {https://doi.org/10.1209/epl/i2001-00248-2} {\bibfield  {journal} {\bibinfo
  {journal} {Europhysics Letters ({EPL})}\ }\textbf {\bibinfo {volume} {54}},\
  \bibinfo {pages} {342} (\bibinfo {year} {2001})}\BibitemShut {NoStop}%
\bibitem [{\citenamefont {Keller}\ \emph {et~al.}(2002)\citenamefont {Keller},
  \citenamefont {Rekveldt},\ and\ \citenamefont {Habicht}}]{Keller2002}%
  \BibitemOpen
  \bibfield  {author} {\bibinfo {author} {\bibfnamefont {T.}~\bibnamefont
  {Keller}}, \bibinfo {author} {\bibfnamefont {M.~T.}\ \bibnamefont
  {Rekveldt}},\ and\ \bibinfo {author} {\bibfnamefont {K.}~\bibnamefont
  {Habicht}},\ }\bibfield  {title} {\bibinfo {title} {Neutron {L}armor
  diffraction measurement of the lattice-spacing spread of pyrolytic
  graphite},\ }\href {https://doi.org/10.1007/s003390101082} {\bibfield
  {journal} {\bibinfo  {journal} {Appl. Phys. A}\ }\textbf {\bibinfo {volume}
  {74}},\ \bibinfo {pages} {S127} (\bibinfo {year} {2002})}\BibitemShut
  {NoStop}%
\bibitem [{\citenamefont {Lorenzo}\ \emph {et~al.}(2012)\citenamefont
  {Lorenzo}, \citenamefont {Joly}, \citenamefont {Mannix},\ and\ \citenamefont
  {Grenier}}]{Lorenzo2012}%
  \BibitemOpen
  \bibfield  {author} {\bibinfo {author} {\bibfnamefont {J.~E.}\ \bibnamefont
  {Lorenzo}}, \bibinfo {author} {\bibfnamefont {Y.}~\bibnamefont {Joly}},
  \bibinfo {author} {\bibfnamefont {D.}~\bibnamefont {Mannix}},\ and\ \bibinfo
  {author} {\bibfnamefont {S.}~\bibnamefont {Grenier}},\ }\bibfield  {title}
  {\bibinfo {title} {Charge order as seen by resonant (elastic) {X}-ray
  scattering},\ }\href {https://doi.org/10.1140/epjst/e2012-01612-5} {\bibfield
   {journal} {\bibinfo  {journal} {Europhys. J. Special Topics}\ }\textbf
  {\bibinfo {volume} {208}},\ \bibinfo {pages} {121} (\bibinfo {year}
  {2012})}\BibitemShut {NoStop}%
\bibitem [{\citenamefont {Fink}\ \emph {et~al.}(2013)\citenamefont {Fink},
  \citenamefont {Schierle}, \citenamefont {Weschke},\ and\ \citenamefont
  {Geck}}]{Fink2013}%
  \BibitemOpen
  \bibfield  {author} {\bibinfo {author} {\bibfnamefont {J.}~\bibnamefont
  {Fink}}, \bibinfo {author} {\bibfnamefont {E.}~\bibnamefont {Schierle}},
  \bibinfo {author} {\bibfnamefont {E.}~\bibnamefont {Weschke}},\ and\ \bibinfo
  {author} {\bibfnamefont {J.}~\bibnamefont {Geck}},\ }\bibfield  {title}
  {\bibinfo {title} {Resonant elastic soft x-ray scattering},\ }\href
  {https://doi.org/10.1088/0034-4885/76/5/056502} {\bibfield  {journal}
  {\bibinfo  {journal} {Reports on Progress in Physics}\ }\textbf {\bibinfo
  {volume} {76}},\ \bibinfo {pages} {056502} (\bibinfo {year} {2013})},\
  \Eprint {https://arxiv.org/abs/1210.5387} {arXiv:1210.5387
  [cond-mat.mtrl-sci]} \BibitemShut {NoStop}%
\bibitem [{\citenamefont {Walz}(2002)}]{Walz2002}%
  \BibitemOpen
  \bibfield  {author} {\bibinfo {author} {\bibfnamefont {F.}~\bibnamefont
  {Walz}},\ }\bibfield  {title} {\bibinfo {title} {The {V}erwey transition---a
  topical review},\ }\href {https://doi.org/10.1088/0953-8984/14/12/203}
  {\bibfield  {journal} {\bibinfo  {journal} {Journal of Physics: Condensed
  Matter}\ }\textbf {\bibinfo {volume} {14}},\ \bibinfo {pages} {R285}
  (\bibinfo {year} {2002})}\BibitemShut {NoStop}%
\bibitem [{\citenamefont {Attfield}(2006)}]{Attfield2006}%
  \BibitemOpen
  \bibfield  {author} {\bibinfo {author} {\bibfnamefont {J.~P.}\ \bibnamefont
  {Attfield}},\ }\bibfield  {title} {\bibinfo {title} {Charge ordering in
  transition metal oxides},\ }\href
  {https://doi.org/10.1016/j.solidstatesciences.2005.02.011} {\bibfield
  {journal} {\bibinfo  {journal} {Solid State Sciences}\ }\textbf {\bibinfo
  {volume} {8}},\ \bibinfo {pages} {861} (\bibinfo {year} {2006})}\BibitemShut
  {NoStop}%
\bibitem [{\citenamefont {Ishihara}(2010)}]{Ishihara2010}%
  \BibitemOpen
  \bibfield  {author} {\bibinfo {author} {\bibfnamefont {S.}~\bibnamefont
  {Ishihara}},\ }\bibfield  {title} {\bibinfo {title} {Electronic
  ferroelectricity and frustration},\ }\href
  {https://doi.org/10.1143/JPSJ.79.011010} {\bibfield  {journal} {\bibinfo
  {journal} {Journal of the Physical Society of Japan}\ }\textbf {\bibinfo
  {volume} {79}},\ \bibinfo {pages} {011010} (\bibinfo {year} {2010})},\
  \Eprint {https://arxiv.org/abs/0912.4083} {arXiv:0912.4083 [cond-mat.str-el]}
  \BibitemShut {NoStop}%
\bibitem [{\citenamefont {Ikeda}\ \emph {et~al.}(2015)\citenamefont {Ikeda},
  \citenamefont {Nagata}, \citenamefont {Kano},\ and\ \citenamefont
  {Mori}}]{Ikeda2015}%
  \BibitemOpen
  \bibfield  {author} {\bibinfo {author} {\bibfnamefont {N.}~\bibnamefont
  {Ikeda}}, \bibinfo {author} {\bibfnamefont {T.}~\bibnamefont {Nagata}},
  \bibinfo {author} {\bibfnamefont {J.}~\bibnamefont {Kano}},\ and\ \bibinfo
  {author} {\bibfnamefont {S.}~\bibnamefont {Mori}},\ }\bibfield  {title}
  {\bibinfo {title} {Present status of the experimental aspect of
  {$R$Fe}$_2${O}$_4$ study},\ }\href
  {https://doi.org/10.1088/0953-8984/27/5/053201} {\bibfield  {journal}
  {\bibinfo  {journal} {Journal of Physics: Condensed Matter}\ }\textbf
  {\bibinfo {volume} {27}},\ \bibinfo {pages} {053201} (\bibinfo {year}
  {2015})}\BibitemShut {NoStop}%
\bibitem [{\citenamefont {Jiang}\ \emph {et~al.}(2014)\citenamefont {Jiang},
  \citenamefont {Zhou},\ and\ \citenamefont {Wang}}]{Jiang2014}%
  \BibitemOpen
  \bibfield  {author} {\bibinfo {author} {\bibfnamefont {K.}~\bibnamefont
  {Jiang}}, \bibinfo {author} {\bibfnamefont {S.}~\bibnamefont {Zhou}},\ and\
  \bibinfo {author} {\bibfnamefont {Z.}~\bibnamefont {Wang}},\ }\bibfield
  {title} {\bibinfo {title} {Textured electronic states of the
  triangular-lattice {H}ubbard model and {Na}$_x${CoO}$_2$},\ }\href
  {https://doi.org/10.1103/PhysRevB.90.165135} {\bibfield  {journal} {\bibinfo
  {journal} {Phys. Rev. B}\ }\textbf {\bibinfo {volume} {90}},\ \bibinfo
  {pages} {165135} (\bibinfo {year} {2014})},\ \Eprint
  {https://arxiv.org/abs/1309.0518} {arXiv:1309.0518 [cond-mat.str-el]}
  \BibitemShut {NoStop}%
\bibitem [{\citenamefont {Cano-Cort{\'e}s}\ \emph {et~al.}(2011)\citenamefont
  {Cano-Cort{\'e}s}, \citenamefont {Ralko}, \citenamefont {F{\'e}vrier},
  \citenamefont {Merino},\ and\ \citenamefont {Fratini}}]{CanoCortes2011}%
  \BibitemOpen
  \bibfield  {author} {\bibinfo {author} {\bibfnamefont {L.}~\bibnamefont
  {Cano-Cort{\'e}s}}, \bibinfo {author} {\bibfnamefont {A.}~\bibnamefont
  {Ralko}}, \bibinfo {author} {\bibfnamefont {C.}~\bibnamefont {F{\'e}vrier}},
  \bibinfo {author} {\bibfnamefont {J.}~\bibnamefont {Merino}},\ and\ \bibinfo
  {author} {\bibfnamefont {S.}~\bibnamefont {Fratini}},\ }\bibfield  {title}
  {\bibinfo {title} {Geometrical frustration effects on charge-driven quantum
  phase transitions},\ }\href {https://doi.org/10.1103/PhysRevB.84.155115}
  {\bibfield  {journal} {\bibinfo  {journal} {Phys. Rev. B}\ }\textbf {\bibinfo
  {volume} {84}},\ \bibinfo {pages} {155115} (\bibinfo {year} {2011})},\
  \Eprint {https://arxiv.org/abs/1106.4408} {arXiv:1106.4408 [cond-mat.str-el]}
  \BibitemShut {NoStop}%
\bibitem [{\citenamefont {Oike}\ \emph {et~al.}(2015)\citenamefont {Oike},
  \citenamefont {Kagawa}, \citenamefont {Ogawa}, \citenamefont {Ueda},
  \citenamefont {Mori}, \citenamefont {Kawasaki},\ and\ \citenamefont
  {Tokura}}]{Oike2015}%
  \BibitemOpen
  \bibfield  {author} {\bibinfo {author} {\bibfnamefont {H.}~\bibnamefont
  {Oike}}, \bibinfo {author} {\bibfnamefont {F.}~\bibnamefont {Kagawa}},
  \bibinfo {author} {\bibfnamefont {N.}~\bibnamefont {Ogawa}}, \bibinfo
  {author} {\bibfnamefont {A.}~\bibnamefont {Ueda}}, \bibinfo {author}
  {\bibfnamefont {H.}~\bibnamefont {Mori}}, \bibinfo {author} {\bibfnamefont
  {M.}~\bibnamefont {Kawasaki}},\ and\ \bibinfo {author} {\bibfnamefont
  {Y.}~\bibnamefont {Tokura}},\ }\bibfield  {title} {\bibinfo {title}
  {Phase-change memory function of correlated electrons in organic
  conductors},\ }\href {https://doi.org/10.1103/PhysRevB.91.041101} {\bibfield
  {journal} {\bibinfo  {journal} {Phys. Rev. B}\ }\textbf {\bibinfo {volume}
  {91}},\ \bibinfo {pages} {041101} (\bibinfo {year} {2015})},\ \Eprint
  {https://arxiv.org/abs/1501.02873} {arXiv:1501.02873 [cond-mat.str-el]}
  \BibitemShut {NoStop}%
\bibitem [{\citenamefont {Borzi}\ \emph {et~al.}(2007)\citenamefont {Borzi},
  \citenamefont {Grigera}, \citenamefont {Farrell}, \citenamefont {Perry},
  \citenamefont {Lister}, \citenamefont {Lee}, \citenamefont {Tennant},
  \citenamefont {Maeno},\ and\ \citenamefont {Mackenzie}}]{Borzi2007}%
  \BibitemOpen
  \bibfield  {author} {\bibinfo {author} {\bibfnamefont {R.~A.}\ \bibnamefont
  {Borzi}}, \bibinfo {author} {\bibfnamefont {S.~A.}\ \bibnamefont {Grigera}},
  \bibinfo {author} {\bibfnamefont {J.}~\bibnamefont {Farrell}}, \bibinfo
  {author} {\bibfnamefont {R.~S.}\ \bibnamefont {Perry}}, \bibinfo {author}
  {\bibfnamefont {S.~J.~S.}\ \bibnamefont {Lister}}, \bibinfo {author}
  {\bibfnamefont {S.~L.}\ \bibnamefont {Lee}}, \bibinfo {author} {\bibfnamefont
  {D.~A.}\ \bibnamefont {Tennant}}, \bibinfo {author} {\bibfnamefont
  {Y.}~\bibnamefont {Maeno}},\ and\ \bibinfo {author} {\bibfnamefont {A.~P.}\
  \bibnamefont {Mackenzie}},\ }\bibfield  {title} {\bibinfo {title} {Formation
  of a nematic fluid at high fields in {Sr$_3$Ru$_2$O$_7$}},\ }\href
  {https://doi.org/10.1126/science.1134796} {\bibfield  {journal} {\bibinfo
  {journal} {Science}\ }\textbf {\bibinfo {volume} {315}},\ \bibinfo {pages}
  {214} (\bibinfo {year} {2007})},\ \Eprint
  {https://arxiv.org/abs/cond-mat/0612599} {arXiv:cond-mat/0612599
  [cond-mat.str-el]} \BibitemShut {NoStop}%
\bibitem [{\citenamefont {Kimura}\ \emph {et~al.}(1996)\citenamefont {Kimura},
  \citenamefont {Tomioka}, \citenamefont {Kuwahara}, \citenamefont {Asamitsu},
  \citenamefont {Tamura},\ and\ \citenamefont {Tokura}}]{Kimura1996}%
  \BibitemOpen
  \bibfield  {author} {\bibinfo {author} {\bibfnamefont {T.}~\bibnamefont
  {Kimura}}, \bibinfo {author} {\bibfnamefont {Y.}~\bibnamefont {Tomioka}},
  \bibinfo {author} {\bibfnamefont {H.}~\bibnamefont {Kuwahara}}, \bibinfo
  {author} {\bibfnamefont {A.}~\bibnamefont {Asamitsu}}, \bibinfo {author}
  {\bibfnamefont {M.}~\bibnamefont {Tamura}},\ and\ \bibinfo {author}
  {\bibfnamefont {Y.}~\bibnamefont {Tokura}},\ }\bibfield  {title} {\bibinfo
  {title} {Interplane tunneling magnetoresistance in a layered manganite
  crystal},\ }\href {https://doi.org/10.1126/science.274.5293.1698} {\bibfield
  {journal} {\bibinfo  {journal} {Science}\ }\textbf {\bibinfo {volume}
  {274}},\ \bibinfo {pages} {1698} (\bibinfo {year} {1996})}\BibitemShut
  {NoStop}%
\bibitem [{\citenamefont {Cava}\ \emph {et~al.}(1990)\citenamefont {Cava},
  \citenamefont {Batlogg}, \citenamefont {van Dover}, \citenamefont
  {Krajewski}, \citenamefont {Waszczak}, \citenamefont {Fleming}, \citenamefont
  {{Peck Jr}}, \citenamefont {{Rupp Jr}}, \citenamefont {Marsh}, \citenamefont
  {James},\ and\ \citenamefont {Schneemeyer}}]{Cava1990}%
  \BibitemOpen
  \bibfield  {author} {\bibinfo {author} {\bibfnamefont {R.~J.}\ \bibnamefont
  {Cava}}, \bibinfo {author} {\bibfnamefont {B.}~\bibnamefont {Batlogg}},
  \bibinfo {author} {\bibfnamefont {R.~B.}\ \bibnamefont {van Dover}}, \bibinfo
  {author} {\bibfnamefont {J.~J.}\ \bibnamefont {Krajewski}}, \bibinfo {author}
  {\bibfnamefont {J.~V.}\ \bibnamefont {Waszczak}}, \bibinfo {author}
  {\bibfnamefont {R.~M.}\ \bibnamefont {Fleming}}, \bibinfo {author}
  {\bibfnamefont {W.~F.}\ \bibnamefont {{Peck Jr}}}, \bibinfo {author}
  {\bibfnamefont {L.~W.}\ \bibnamefont {{Rupp Jr}}}, \bibinfo {author}
  {\bibfnamefont {P.}~\bibnamefont {Marsh}}, \bibinfo {author} {\bibfnamefont
  {A.~C. W.~P.}\ \bibnamefont {James}},\ and\ \bibinfo {author} {\bibfnamefont
  {L.~F.}\ \bibnamefont {Schneemeyer}},\ }\bibfield  {title} {\bibinfo {title}
  {Superconductivity at 60\,{K} in {La$_{2-x}$Sr$_x$CaCu$_2$O$_6$}: the
  simplest double-layer cuprate},\ }\href {https://doi.org/10.1038/345602a0}
  {\bibfield  {journal} {\bibinfo  {journal} {Nature}\ }\textbf {\bibinfo
  {volume} {345}},\ \bibinfo {pages} {602} (\bibinfo {year}
  {1990})}\BibitemShut {NoStop}%
\bibitem [{\citenamefont {Kim}\ \emph {et~al.}(2014)\citenamefont {Kim},
  \citenamefont {Jain}, \citenamefont {Reehuis}, \citenamefont {Khaliullin},
  \citenamefont {Peets}, \citenamefont {Ulrich}, \citenamefont {Park},
  \citenamefont {Faulhaber}, \citenamefont {Hoser}, \citenamefont {Walker},
  \citenamefont {Adroja}, \citenamefont {Walters}, \citenamefont {Inosov},
  \citenamefont {Maljuk},\ and\ \citenamefont {Keimer}}]{Kim2014}%
  \BibitemOpen
  \bibfield  {author} {\bibinfo {author} {\bibfnamefont {J.-H.}\ \bibnamefont
  {Kim}}, \bibinfo {author} {\bibfnamefont {A.}~\bibnamefont {Jain}}, \bibinfo
  {author} {\bibfnamefont {M.}~\bibnamefont {Reehuis}}, \bibinfo {author}
  {\bibfnamefont {G.}~\bibnamefont {Khaliullin}}, \bibinfo {author}
  {\bibfnamefont {D.~C.}\ \bibnamefont {Peets}}, \bibinfo {author}
  {\bibfnamefont {C.}~\bibnamefont {Ulrich}}, \bibinfo {author} {\bibfnamefont
  {J.~T.}\ \bibnamefont {Park}}, \bibinfo {author} {\bibfnamefont
  {E.}~\bibnamefont {Faulhaber}}, \bibinfo {author} {\bibfnamefont
  {A.}~\bibnamefont {Hoser}}, \bibinfo {author} {\bibfnamefont {H.~C.}\
  \bibnamefont {Walker}}, \bibinfo {author} {\bibfnamefont {D.~T.}\
  \bibnamefont {Adroja}}, \bibinfo {author} {\bibfnamefont {A.~C.}\
  \bibnamefont {Walters}}, \bibinfo {author} {\bibfnamefont {D.~S.}\
  \bibnamefont {Inosov}}, \bibinfo {author} {\bibfnamefont {A.}~\bibnamefont
  {Maljuk}},\ and\ \bibinfo {author} {\bibfnamefont {B.}~\bibnamefont
  {Keimer}},\ }\bibfield  {title} {\bibinfo {title} {Competing exchange
  interactions on the verge of a metal-insulator transition in the
  two-dimensional spiral magnet {Sr}$_3${Fe}$_2${O}$_7$},\ }\href
  {https://doi.org/10.1103/PhysRevLett.113.147206} {\bibfield  {journal}
  {\bibinfo  {journal} {Phys. Rev. Lett.}\ }\textbf {\bibinfo {volume} {113}},\
  \bibinfo {pages} {147206} (\bibinfo {year} {2014})},\ \Eprint
  {https://arxiv.org/abs/1409.5205} {arXiv:1409.5205 [cond-mat.str-el]}
  \BibitemShut {NoStop}%
\bibitem [{\citenamefont {Mazin}\ \emph {et~al.}(2007)\citenamefont {Mazin},
  \citenamefont {Khomskii}, \citenamefont {Lengsdorf}, \citenamefont {Alonso},
  \citenamefont {Marshall}, \citenamefont {Ibberson}, \citenamefont
  {Podlesnyak}, \citenamefont {Mart\'{\i}nez-Lope},\ and\ \citenamefont
  {Abd-Elmeguid}}]{Mazin2007}%
  \BibitemOpen
  \bibfield  {author} {\bibinfo {author} {\bibfnamefont {I.~I.}\ \bibnamefont
  {Mazin}}, \bibinfo {author} {\bibfnamefont {D.~I.}\ \bibnamefont {Khomskii}},
  \bibinfo {author} {\bibfnamefont {R.}~\bibnamefont {Lengsdorf}}, \bibinfo
  {author} {\bibfnamefont {J.~A.}\ \bibnamefont {Alonso}}, \bibinfo {author}
  {\bibfnamefont {W.~G.}\ \bibnamefont {Marshall}}, \bibinfo {author}
  {\bibfnamefont {R.~M.}\ \bibnamefont {Ibberson}}, \bibinfo {author}
  {\bibfnamefont {A.}~\bibnamefont {Podlesnyak}}, \bibinfo {author}
  {\bibfnamefont {M.~J.}\ \bibnamefont {Mart\'{\i}nez-Lope}},\ and\ \bibinfo
  {author} {\bibfnamefont {M.~M.}\ \bibnamefont {Abd-Elmeguid}},\ }\bibfield
  {title} {\bibinfo {title} {Charge ordering as alternative to {J}ahn-{T}eller
  distortion},\ }\href {https://doi.org/10.1103/PhysRevLett.98.176406}
  {\bibfield  {journal} {\bibinfo  {journal} {Phys. Rev. Lett.}\ }\textbf
  {\bibinfo {volume} {98}},\ \bibinfo {pages} {176406} (\bibinfo {year}
  {2007})}\BibitemShut {NoStop}%
\bibitem [{Kim()}]{KimPRL2019Supp}%
  \BibitemOpen
  \href@noop {} {}\bibinfo {note} {See Supplemental Material in the Appendix
  below, including
  Refs.~\onlinecite{Hall1995,ITC2006,Fullprof2,Mosswinn,Takano1977,Demazeau1981},
  for additional experimental details, crystallographic parameters, and
  M{\"o}ssbauer results. CIF files of our crystal structure refinements are
  available as ArXiv ancillary files.}\BibitemShut {Stop}%
\bibitem [{\citenamefont {Maljuk}\ \emph {et~al.}(2004)\citenamefont {Maljuk},
  \citenamefont {Strempfer}, \citenamefont {Ulrich}, \citenamefont {Sofin},
  \citenamefont {Capogna}, \citenamefont {Lin},\ and\ \citenamefont
  {Keimer}}]{Maljuk2004}%
  \BibitemOpen
  \bibfield  {author} {\bibinfo {author} {\bibfnamefont {A.}~\bibnamefont
  {Maljuk}}, \bibinfo {author} {\bibfnamefont {J.}~\bibnamefont {Strempfer}},
  \bibinfo {author} {\bibfnamefont {C.}~\bibnamefont {Ulrich}}, \bibinfo
  {author} {\bibfnamefont {M.}~\bibnamefont {Sofin}}, \bibinfo {author}
  {\bibfnamefont {L.}~\bibnamefont {Capogna}}, \bibinfo {author} {\bibfnamefont
  {C.}~\bibnamefont {Lin}},\ and\ \bibinfo {author} {\bibfnamefont
  {B.}~\bibnamefont {Keimer}},\ }\bibfield  {title} {\bibinfo {title} {Growth
  of {Sr$_3$Fe$_2$O$_{7-x}$} single crystals by the floating zone method},\
  }\href {https://doi.org/10.1016/j.jcrysgro.2004.07.091} {\bibfield  {journal}
  {\bibinfo  {journal} {Journal of Crystal Growth}\ }\textbf {\bibinfo {volume}
  {273}},\ \bibinfo {pages} {207} (\bibinfo {year} {2004})}\BibitemShut
  {NoStop}%
\bibitem [{\citenamefont {Bocquet}\ \emph {et~al.}(1992)\citenamefont
  {Bocquet}, \citenamefont {Fujimori}, \citenamefont {Mizokawa}, \citenamefont
  {Saitoh}, \citenamefont {Namatame}, \citenamefont {Suga}, \citenamefont
  {Kimizuka}, \citenamefont {Takeda},\ and\ \citenamefont
  {Takano}}]{Bocquet1992}%
  \BibitemOpen
  \bibfield  {author} {\bibinfo {author} {\bibfnamefont {A.~E.}\ \bibnamefont
  {Bocquet}}, \bibinfo {author} {\bibfnamefont {A.}~\bibnamefont {Fujimori}},
  \bibinfo {author} {\bibfnamefont {T.}~\bibnamefont {Mizokawa}}, \bibinfo
  {author} {\bibfnamefont {T.}~\bibnamefont {Saitoh}}, \bibinfo {author}
  {\bibfnamefont {H.}~\bibnamefont {Namatame}}, \bibinfo {author}
  {\bibfnamefont {S.}~\bibnamefont {Suga}}, \bibinfo {author} {\bibfnamefont
  {N.}~\bibnamefont {Kimizuka}}, \bibinfo {author} {\bibfnamefont
  {Y.}~\bibnamefont {Takeda}},\ and\ \bibinfo {author} {\bibfnamefont
  {M.}~\bibnamefont {Takano}},\ }\bibfield  {title} {\bibinfo {title}
  {Electronic structure of {SrFe$^{4+}$O$_3$} and related {Fe} perovskite
  oxides},\ }\href {https://doi.org/10.1103/PhysRevB.45.1561} {\bibfield
  {journal} {\bibinfo  {journal} {Phys. Rev. B}\ }\textbf {\bibinfo {volume}
  {45}},\ \bibinfo {pages} {1561} (\bibinfo {year} {1992})}\BibitemShut
  {NoStop}%
\bibitem [{\citenamefont {Green}\ \emph {et~al.}(2016)\citenamefont {Green},
  \citenamefont {Haverkort},\ and\ \citenamefont {Sawatzky}}]{Green2016}%
  \BibitemOpen
  \bibfield  {author} {\bibinfo {author} {\bibfnamefont {R.~J.}\ \bibnamefont
  {Green}}, \bibinfo {author} {\bibfnamefont {M.~W.}\ \bibnamefont
  {Haverkort}},\ and\ \bibinfo {author} {\bibfnamefont {G.~A.}\ \bibnamefont
  {Sawatzky}},\ }\bibfield  {title} {\bibinfo {title} {Bond disproportionation
  and dynamical charge fluctuations in the perovskite rare-earth nickelates},\
  }\href {https://doi.org/10.1103/PhysRevB.94.195127} {\bibfield  {journal}
  {\bibinfo  {journal} {Phys. Rev. B}\ }\textbf {\bibinfo {volume} {94}},\
  \bibinfo {pages} {195127} (\bibinfo {year} {2016})},\ \Eprint
  {https://arxiv.org/abs/1608.01645} {arXiv:1608.01645 [cond-mat.str-el]}
  \BibitemShut {NoStop}%
\bibitem [{\citenamefont {Keller}\ and\ \citenamefont
  {Keimer}(2015)}]{TRISP2015}%
  \BibitemOpen
  \bibfield  {author} {\bibinfo {author} {\bibfnamefont {T.}~\bibnamefont
  {Keller}}\ and\ \bibinfo {author} {\bibfnamefont {B.}~\bibnamefont
  {Keimer}},\ }\bibfield  {title} {\bibinfo {title} {{TRISP}: Three axes spin
  echo spectrometer},\ }\href {https://doi.org/10.17815/jlsrf-1-41} {\bibfield
  {journal} {\bibinfo  {journal} {Journal of large-scale research facilities}\
  }\textbf {\bibinfo {volume} {1}},\ \bibinfo {pages} {A37} (\bibinfo {year}
  {2015})}\BibitemShut {NoStop}%
\bibitem [{\citenamefont {Woodward}\ \emph {et~al.}(2000)\citenamefont
  {Woodward}, \citenamefont {Cox}, \citenamefont {Moshopoulou}, \citenamefont
  {Sleight},\ and\ \citenamefont {Morimoto}}]{Woodward2000}%
  \BibitemOpen
  \bibfield  {author} {\bibinfo {author} {\bibfnamefont {P.~M.}\ \bibnamefont
  {Woodward}}, \bibinfo {author} {\bibfnamefont {D.~E.}\ \bibnamefont {Cox}},
  \bibinfo {author} {\bibfnamefont {E.}~\bibnamefont {Moshopoulou}}, \bibinfo
  {author} {\bibfnamefont {A.~W.}\ \bibnamefont {Sleight}},\ and\ \bibinfo
  {author} {\bibfnamefont {S.}~\bibnamefont {Morimoto}},\ }\bibfield  {title}
  {\bibinfo {title} {Structural studies of charge disproportionation and
  magnetic order in {CaFeO}$_3$},\ }\href
  {https://doi.org/10.1103/PhysRevB.62.844} {\bibfield  {journal} {\bibinfo
  {journal} {Phys. Rev. B}\ }\textbf {\bibinfo {volume} {62}},\ \bibinfo
  {pages} {844} (\bibinfo {year} {2000})}\BibitemShut {NoStop}%
\bibitem [{\citenamefont {Shaked}\ \emph {et~al.}(2000)\citenamefont {Shaked},
  \citenamefont {Jorgensen}, \citenamefont {Chmaissem}, \citenamefont {Ikeda},\
  and\ \citenamefont {Maeno}}]{Shaked2000}%
  \BibitemOpen
  \bibfield  {author} {\bibinfo {author} {\bibfnamefont {H.}~\bibnamefont
  {Shaked}}, \bibinfo {author} {\bibfnamefont {J.~D.}\ \bibnamefont
  {Jorgensen}}, \bibinfo {author} {\bibfnamefont {O.}~\bibnamefont
  {Chmaissem}}, \bibinfo {author} {\bibfnamefont {S.}~\bibnamefont {Ikeda}},\
  and\ \bibinfo {author} {\bibfnamefont {Y.}~\bibnamefont {Maeno}},\ }\bibfield
   {title} {\bibinfo {title} {Neutron diffraction study of the structural
  distortions in {Sr$_3$Ru$_2$O$_7$}},\ }\href
  {https://doi.org/10.1006/jssc.2000.8796} {\bibfield  {journal} {\bibinfo
  {journal} {J.\ Solid State Chem.}\ }\textbf {\bibinfo {volume} {154}},\
  \bibinfo {pages} {361} (\bibinfo {year} {2000})}\BibitemShut {NoStop}%
\bibitem [{\citenamefont {Kiyanagi}\ \emph {et~al.}(2004)\citenamefont
  {Kiyanagi}, \citenamefont {Tsuda}, \citenamefont {Aso}, \citenamefont
  {Kimura}, \citenamefont {Noda}, \citenamefont {Yoshida}, \citenamefont
  {Ikeda},\ and\ \citenamefont {Uwatoko}}]{Kiyanagi2004}%
  \BibitemOpen
  \bibfield  {author} {\bibinfo {author} {\bibfnamefont {R.}~\bibnamefont
  {Kiyanagi}}, \bibinfo {author} {\bibfnamefont {K.}~\bibnamefont {Tsuda}},
  \bibinfo {author} {\bibfnamefont {N.}~\bibnamefont {Aso}}, \bibinfo {author}
  {\bibfnamefont {H.}~\bibnamefont {Kimura}}, \bibinfo {author} {\bibfnamefont
  {Y.}~\bibnamefont {Noda}}, \bibinfo {author} {\bibfnamefont {Y.}~\bibnamefont
  {Yoshida}}, \bibinfo {author} {\bibfnamefont {S.-I.}\ \bibnamefont {Ikeda}},\
  and\ \bibinfo {author} {\bibfnamefont {Y.}~\bibnamefont {Uwatoko}},\
  }\bibfield  {title} {\bibinfo {title} {Investigation of the structure of
  single crystal {Sr$_3$Ru$_2$O$_7$} by neutron and convergent beam electron
  diffractions},\ }\href {https://doi.org/10.1143/JPSJ.73.639} {\bibfield
  {journal} {\bibinfo  {journal} {J.\ Phys.\ Soc.\ Japan}\ }\textbf {\bibinfo
  {volume} {73}},\ \bibinfo {pages} {639} (\bibinfo {year} {2004})}\BibitemShut
  {NoStop}%
\bibitem [{\citenamefont {Strempfer}\ \emph {et~al.}(2013)\citenamefont
  {Strempfer}, \citenamefont {Francoual}, \citenamefont {Reuther},
  \citenamefont {Shukla}, \citenamefont {Skaugen}, \citenamefont
  {Schulte-Schrepping}, \citenamefont {Krachta},\ and\ \citenamefont
  {Franz}}]{Strempfer2013}%
  \BibitemOpen
  \bibfield  {author} {\bibinfo {author} {\bibfnamefont {J.}~\bibnamefont
  {Strempfer}}, \bibinfo {author} {\bibfnamefont {S.}~\bibnamefont
  {Francoual}}, \bibinfo {author} {\bibfnamefont {D.}~\bibnamefont {Reuther}},
  \bibinfo {author} {\bibfnamefont {D.~K.}\ \bibnamefont {Shukla}}, \bibinfo
  {author} {\bibfnamefont {A.}~\bibnamefont {Skaugen}}, \bibinfo {author}
  {\bibfnamefont {H.}~\bibnamefont {Schulte-Schrepping}}, \bibinfo {author}
  {\bibfnamefont {T.}~\bibnamefont {Krachta}},\ and\ \bibinfo {author}
  {\bibfnamefont {H.}~\bibnamefont {Franz}},\ }\bibfield  {title} {\bibinfo
  {title} {Resonant scattering and diffraction beamline {P09} at {PETRA III}},\
  }\href {https://doi.org/10.1107/S0909049513009011} {\bibfield  {journal}
  {\bibinfo  {journal} {Journal of Synchrotron Radiation}\ }\textbf {\bibinfo
  {volume} {20}},\ \bibinfo {pages} {541} (\bibinfo {year} {2013})}\BibitemShut
  {NoStop}%
\bibitem [{\citenamefont {Lu}\ \emph {et~al.}(2016)\citenamefont {Lu},
  \citenamefont {Frano}, \citenamefont {Bluschke}, \citenamefont {Hepting},
  \citenamefont {Macke}, \citenamefont {Strempfer}, \citenamefont {Wochner},
  \citenamefont {Cristiani}, \citenamefont {Logvenov}, \citenamefont
  {Habermeier}, \citenamefont {Haverkort}, \citenamefont {Keimer},\ and\
  \citenamefont {Benckiser}}]{Lu2016}%
  \BibitemOpen
  \bibfield  {author} {\bibinfo {author} {\bibfnamefont {Y.}~\bibnamefont
  {Lu}}, \bibinfo {author} {\bibfnamefont {A.}~\bibnamefont {Frano}}, \bibinfo
  {author} {\bibfnamefont {M.}~\bibnamefont {Bluschke}}, \bibinfo {author}
  {\bibfnamefont {M.}~\bibnamefont {Hepting}}, \bibinfo {author} {\bibfnamefont
  {S.}~\bibnamefont {Macke}}, \bibinfo {author} {\bibfnamefont
  {J.}~\bibnamefont {Strempfer}}, \bibinfo {author} {\bibfnamefont
  {P.}~\bibnamefont {Wochner}}, \bibinfo {author} {\bibfnamefont
  {G.}~\bibnamefont {Cristiani}}, \bibinfo {author} {\bibfnamefont
  {G.}~\bibnamefont {Logvenov}}, \bibinfo {author} {\bibfnamefont {H.-U.}\
  \bibnamefont {Habermeier}}, \bibinfo {author} {\bibfnamefont {M.~W.}\
  \bibnamefont {Haverkort}}, \bibinfo {author} {\bibfnamefont {B.}~\bibnamefont
  {Keimer}},\ and\ \bibinfo {author} {\bibfnamefont {E.}~\bibnamefont
  {Benckiser}},\ }\bibfield  {title} {\bibinfo {title} {Quantitative
  determination of bond order and lattice distortions in nickel oxide
  heterostructures by resonant x-ray scattering},\ }\href
  {https://doi.org/10.1103/PhysRevB.93.165121} {\bibfield  {journal} {\bibinfo
  {journal} {Phys.\ Rev.\ B}\ }\textbf {\bibinfo {volume} {93}},\ \bibinfo
  {pages} {165121} (\bibinfo {year} {2016})},\ \Eprint
  {https://arxiv.org/abs/1604.07317} {arXiv:1604.07317 [cond-mat.str-el]}
  \BibitemShut {NoStop}%
\bibitem [{\citenamefont {Onoda}\ \emph {et~al.}(2004)\citenamefont {Onoda},
  \citenamefont {Motome},\ and\ \citenamefont {Nagaosa}}]{Onoda2004}%
  \BibitemOpen
  \bibfield  {author} {\bibinfo {author} {\bibfnamefont {S.}~\bibnamefont
  {Onoda}}, \bibinfo {author} {\bibfnamefont {Y.}~\bibnamefont {Motome}},\ and\
  \bibinfo {author} {\bibfnamefont {N.}~\bibnamefont {Nagaosa}},\ }\bibfield
  {title} {\bibinfo {title} {Two-dimensional charge order in layered 2-1-4
  perovskite oxides},\ }\href {https://doi.org/10.1103/PhysRevLett.92.236403}
  {\bibfield  {journal} {\bibinfo  {journal} {Phys. Rev. Lett.}\ }\textbf
  {\bibinfo {volume} {92}},\ \bibinfo {pages} {236403} (\bibinfo {year}
  {2004})},\ \Eprint {https://arxiv.org/abs/cond-mat/0211520}
  {arXiv:cond-mat/0211520 [cond-mat.stat-mech]} \BibitemShut {NoStop}%
\bibitem [{\citenamefont {Okazaki}\ \emph {et~al.}(2011)\citenamefont
  {Okazaki}, \citenamefont {Shibauchi}, \citenamefont {Shi}, \citenamefont
  {Haga}, \citenamefont {Matsuda}, \citenamefont {Yamamoto}, \citenamefont
  {Onuki}, \citenamefont {Ikeda},\ and\ \citenamefont {Matsuda}}]{Okazaki2011}%
  \BibitemOpen
  \bibfield  {author} {\bibinfo {author} {\bibfnamefont {R.}~\bibnamefont
  {Okazaki}}, \bibinfo {author} {\bibfnamefont {T.}~\bibnamefont {Shibauchi}},
  \bibinfo {author} {\bibfnamefont {H.~J.}\ \bibnamefont {Shi}}, \bibinfo
  {author} {\bibfnamefont {Y.}~\bibnamefont {Haga}}, \bibinfo {author}
  {\bibfnamefont {T.~D.}\ \bibnamefont {Matsuda}}, \bibinfo {author}
  {\bibfnamefont {E.}~\bibnamefont {Yamamoto}}, \bibinfo {author}
  {\bibfnamefont {Y.}~\bibnamefont {Onuki}}, \bibinfo {author} {\bibfnamefont
  {H.}~\bibnamefont {Ikeda}},\ and\ \bibinfo {author} {\bibfnamefont
  {Y.}~\bibnamefont {Matsuda}},\ }\bibfield  {title} {\bibinfo {title}
  {Rotational symmetry breaking in the hidden-order phase of {URu$_2$Si$_2$}},\
  }\href {https://doi.org/10.1126/science.1197358} {\bibfield  {journal}
  {\bibinfo  {journal} {Science}\ }\textbf {\bibinfo {volume} {331}},\ \bibinfo
  {pages} {439} (\bibinfo {year} {2011})},\ \Eprint
  {https://arxiv.org/abs/1107.5480} {arXiv:1107.5480 [cond-mat.str-el]}
  \BibitemShut {NoStop}%
\bibitem [{\citenamefont {Hall}\ \emph {et~al.}(1995)\citenamefont {Hall},
  \citenamefont {King},\ and\ \citenamefont {Stewart}}]{Hall1995}%
  \BibitemOpen
  \bibinfo {editor} {\bibfnamefont {S.~R.}\ \bibnamefont {Hall}}, \bibinfo
  {editor} {\bibfnamefont {G.~S.~D.}\ \bibnamefont {King}},\ and\ \bibinfo
  {editor} {\bibfnamefont {J.~M.}\ \bibnamefont {Stewart}},\ eds.,\ \href@noop
  {} {\emph {\bibinfo {title} {The {\sc Xtal3.4} User's Manual}}}\ (\bibinfo
  {publisher} {Lamb Print, University of Western Australia},\ \bibinfo
  {address} {Perth},\ \bibinfo {year} {1995})\BibitemShut {NoStop}%
\bibitem [{\citenamefont {Prince}(2006)}]{ITC2006}%
  \BibitemOpen
  \bibinfo {editor} {\bibfnamefont {E.}~\bibnamefont {Prince}},\ ed.,\ \href
  {https://doi.org/10.1107/97809553602060000103} {\emph {\bibinfo {title}
  {International Tables of Crystallography}}},\ Vol.~\bibinfo {volume} {C}\
  (\bibinfo  {publisher} {International Union of Crystallography},\ \bibinfo
  {year} {2006})\BibitemShut {NoStop}%
\bibitem [{\citenamefont {Rodr{\'i}guez-{C}arvajal}(1993)}]{Fullprof2}%
  \BibitemOpen
  \bibfield  {author} {\bibinfo {author} {\bibfnamefont {J.}~\bibnamefont
  {Rodr{\'i}guez-{C}arvajal}},\ }\bibfield  {title} {\bibinfo {title} {Recent
  advances in magnetic structure determination by neutron powder diffraction},\
  }\href {https://doi.org/10.1016/0921-4526(93)90108-I} {\bibfield  {journal}
  {\bibinfo  {journal} {Physica B: Condensed Matter}\ }\textbf {\bibinfo
  {volume} {192}},\ \bibinfo {pages} {55} (\bibinfo {year} {1993})}\BibitemShut
  {NoStop}%
\bibitem [{\citenamefont {Klencs{\'a}r}(2013)}]{Mosswinn}%
  \BibitemOpen
  \bibfield  {author} {\bibinfo {author} {\bibfnamefont {Z.}~\bibnamefont
  {Klencs{\'a}r}},\ }\bibfield  {title} {\bibinfo {title} {{\sc
  MossWinn}---methodological advances in the field of {M}{\"o}ssbauer data
  analysis},\ }\href {https://doi.org/10.1007/s10751-012-0732-2} {\bibfield
  {journal} {\bibinfo  {journal} {Hyperfine Interactions}\ }\textbf {\bibinfo
  {volume} {217}},\ \bibinfo {pages} {117} (\bibinfo {year}
  {2013})}\BibitemShut {NoStop}%
\bibitem [{\citenamefont {Takano}\ \emph {et~al.}(1977)\citenamefont {Takano},
  \citenamefont {Nakanishi}, \citenamefont {Takeda}, \citenamefont {Naka},\
  and\ \citenamefont {Takada}}]{Takano1977}%
  \BibitemOpen
  \bibfield  {author} {\bibinfo {author} {\bibfnamefont {M.}~\bibnamefont
  {Takano}}, \bibinfo {author} {\bibfnamefont {N.}~\bibnamefont {Nakanishi}},
  \bibinfo {author} {\bibfnamefont {Y.}~\bibnamefont {Takeda}}, \bibinfo
  {author} {\bibfnamefont {S.}~\bibnamefont {Naka}},\ and\ \bibinfo {author}
  {\bibfnamefont {T.}~\bibnamefont {Takada}},\ }\bibfield  {title} {\bibinfo
  {title} {Charge disproportionation in {CaFeO$_3$} studied with the
  {M}{\"o}ssbauer effect},\ }\href
  {https://doi.org/10.1016/0025-5408(77)90104-0} {\bibfield  {journal}
  {\bibinfo  {journal} {Materials Research Bulletin}\ }\textbf {\bibinfo
  {volume} {12}},\ \bibinfo {pages} {923} (\bibinfo {year} {1977})}\BibitemShut
  {NoStop}%
\bibitem [{\citenamefont {Demazeau}\ \emph {et~al.}(1981)\citenamefont
  {Demazeau}, \citenamefont {Buffat}, \citenamefont {M{\'e}nil}, \citenamefont
  {Fourn{\`e}s}, \citenamefont {Pouchard}, \citenamefont {Dance}, \citenamefont
  {Fabritchnyi},\ and\ \citenamefont {Hagenmuller}}]{Demazeau1981}%
  \BibitemOpen
  \bibfield  {author} {\bibinfo {author} {\bibfnamefont {G.}~\bibnamefont
  {Demazeau}}, \bibinfo {author} {\bibfnamefont {B.}~\bibnamefont {Buffat}},
  \bibinfo {author} {\bibfnamefont {F.}~\bibnamefont {M{\'e}nil}}, \bibinfo
  {author} {\bibfnamefont {L.}~\bibnamefont {Fourn{\`e}s}}, \bibinfo {author}
  {\bibfnamefont {M.}~\bibnamefont {Pouchard}}, \bibinfo {author}
  {\bibfnamefont {J.~M.}\ \bibnamefont {Dance}}, \bibinfo {author}
  {\bibfnamefont {P.}~\bibnamefont {Fabritchnyi}},\ and\ \bibinfo {author}
  {\bibfnamefont {P.}~\bibnamefont {Hagenmuller}},\ }\bibfield  {title}
  {\bibinfo {title} {Characterization of six-coordinated iron {(V)} in an oxide
  lattice},\ }\href {https://doi.org/10.1016/0025-5408(81)90067-2} {\bibfield
  {journal} {\bibinfo  {journal} {Materials Research Bulletin}\ }\textbf
  {\bibinfo {volume} {16}},\ \bibinfo {pages} {1465} (\bibinfo {year}
  {1981})}\BibitemShut {NoStop}%
\end{thebibliography}%
